\documentclass[conf]{new-aiaa}
\usepackage[utf8]{inputenc}
\usepackage{textcomp}
\usepackage{xcolor,soul}
\sethlcolor{green}
\usepackage{bm}

\usepackage{algorithm}
\usepackage[noend]{algpseudocode}

\usepackage{graphicx}
\usepackage{subfigure}  
\usepackage{amsmath}
\usepackage[version=4]{mhchem}
\usepackage{siunitx}
\usepackage{longtable,tabularx}
\title{Parameter Estimation for RANS Models Using Approximate Bayesian Computation}

\author{Olga A. Doronina$^1$\footnote{Research Assistant, Department of Mechanical Engineering, UCB 427, Boulder, CO, AIAA Student Member.}, Scott M. Murman$^2$\footnote{Aerospace Engineer, NASA Ames Research Center, Moffett Field, CA, USA, AIAA Member.}, Peter E. Hamlington$^1$\footnote{Associate Professor, Department of Mechanical Engineering, UCB 427, Boulder, CO, AIAA Member.}}
\affil{$^1$ Paul M. Rady Department of Mechanical Engineering, University of Colorado, Boulder, CO, USA, 80309}
\affil{$^2$ NASA Ames Research Center, Moffett Field, CA, USA, 94035}

\begin{document}

\maketitle

\begin{abstract}

We use approximate Bayesian computation (ABC) to estimate unknown parameter values, as well as their uncertainties, in Reynolds-averaged Navier-Stokes (RANS) simulations of turbulent flows. The ABC method approximates posterior distributions of model parameters, but does not require the direct computation, or estimation, of a likelihood function. Compared to full Bayesian analyses, ABC thus provides a faster and more flexible parameter estimation for complex models and a wide range of reference data. In this paper, we describe the ABC approach, including the use of a calibration step, adaptive proposal, and Markov chain Monte Carlo (MCMC) technique to accelerate the parameter estimation, resulting in an improved ABC approach, denoted ABC-IMCMC. As a test of the classic ABC rejection algorithm, we estimate parameters in a nonequilibrium RANS model using reference data from direct numerical simulations of periodically sheared homogeneous turbulence. We then demonstrate the use of ABC-IMCMC to estimate parameters in the Menter shear-stress-transport (SST) model using experimental reference data for an axisymmetric transonic bump. We show that the accuracy of the SST model for this test case can be improved using ABC-IMCMC, indicating that ABC-IMCMC is a promising method for the calibration of RANS models using a wide range of reference data. 
\end{abstract}

\section{Introduction}
Despite ongoing advances in the availability of high-performance computing resources, most industrial and engineering fluid flows are still simulated using Reynolds-averaged Navier-Stokes (RANS) approaches. Although progress has been made in recent years on the development of computationally efficient large-eddy simulations (LES) and hybrid RANS/LES approaches, RANS simulations are likely to remain common in engineering practice for many years to come, primarily because of their substantially lower computational cost compared to LES. Techniques such as optimization and uncertainty quantification are also becoming increasingly important in the engineering design process, but such techniques often require thousands of simulations (or more), and RANS simulations remain the only viable option for such large numbers of calculations.

The primary difficulty with RANS simulations is the requirement that a physically accurate, computationally stable model be provided for the unclosed Reynolds stresses that appear in the ensemble-averaged governing equations. A wide variety of RANS closure models have been suggested over the years (see \citet{speziale1998} for a review) but, despite all efforts to create a robust and accurate model, no universal turbulence model exists. Moreover, essentially all RANS models rely on empirical coefficients that must be calibrated for different flows and geometries.

Given this situation, the development of improved methods for inferring model parameters from available experimental, or higher fidelity computational, data is an active area of research~\cite{xiao_quantification_2019}. In particular, the inverse statistical problem that is the core of model parameter calibration can be solved using either deterministic or probabilistic approaches. In the deterministic approach, model parameters are determined through an optimization technique by minimizing the error between model and some reference data. Probabilistic approaches such as Bayesian methods, by contrast, are attractive because they provide not only model parameter values but also their uncertainties. \citet{oberkampf2000validation,oberkampf2002verification} have emphasized the importance of such uncertainties in model parameter calibration, particularly when parameter estimation is performed using uncertain reference data and inherently imperfect models.

\citet{cheung_bayesian_2011} were the first to apply a Bayesian inference method to calibrate the Spalart-Allmaras turbulence model using velocity and wall-shear stress experimental data for wall-bounded, incompressible, turbulent flows. \citet{oliver_bayesian_2011} extended this work by adding more models and stochastic extensions with direct numerical simulation (DNS) channel flow data. Ray and co-authors \cite{ray_bayesian_2014,lefantzi2015estimation, ray_bayesian_2016, ray_learning_2018, ray2018robust} used a similar approach to calibrate RANS model parameters in a jet in crossflow. \citet{zhang_efficient_2019} combined a high-dimensional model representation technique and the Gaussian-process machine-learning method to construct a surrogate model, thereby making the Bayesian inference process more affordable. As a demonstration, they calibrated various parameters in the Menter shear-stress-transport (SST) model, using reference data comprised of surface drag measurements and velocity profiles for hypersonic turbulent flows over flat plates. 

Despite the overall success of these prior approaches, however, solving the full Bayes' problem requires knowledge of the likelihood function, which can be difficult, and/or costly, to compute. In many cases, this likelihood function is approximated by a Gaussian distribution, which does not always reflect the actual likelihood function, or is obtained using a surrogate model that is trained using available RANS simulations~\cite{ray_bayesian_2016, ray_learning_2018, edeling_bayesian_2014}. Here, we outline an alternative Bayesian approach based on the combination of approximate Bayesian computation (ABC) and Markov chain Monte Carlo (MCMC) sampling to determine unknown model parameters and their uncertainties. The ABC approach allows us to estimate the posterior distribution of model parameters, given some reference data, without knowledge of the likelihood function; as such, ABC is often referred to as a ``likelihood-free'' Bayesian method. 

In the present study, we verify and validate the core ABC rejection algorithm through the estimation of parameters in the same nonequilibrium turbulence model examined in Ref.~\cite{doronina2019}, using reference data from the model itself (i.e., verification) and from DNS of periodically sheared homogeneous turbulence \cite{yu_direct_2006} (i.e., validation). We then demonstrate the use of an improved ABC approach with a calibration step, adaptive proposal, and MCMC sampling for the calibration of parameters in the Menter SST model. We show that the efficiency of the improved approach, termed ABC-IMCMC, permits the estimation of parameters using full computational fluid dynamics (CFD) simulations of inhomogeneous flows. In particular, we use ABC-IMCMC to estimate parameters in the Menter SST model given experimental reference data for an axisymmetric transonic bump \cite{bachalo1986transonic}, and using CFD model evaluations in \verb OVERFLOW ~\cite{nichols2006solver}. Recently, \citet{schaefer_uncertainty_2017} and \citet{zhao_uncertainty_2019} performed an uncertainty and sensitivity analysis of the Menter SST model for different turbulent flows, and we perform a related calibration study prior to initiating the Markov chains in order to determine which of the model parameters to include in the estimation procedure.  

In the following, we describe the ABC-IMCMC method in Section~\ref{sec:ABC}. In Section~\ref{sec:nonequil_result}, we describe the verification and validation tests for the nonequilibrium RANS model using the classic ABC rejection algorithm. In Section~\ref{sec:overflow}, we then present results from the estimation of parameters in the Menter SST RANS model using ABC-IMCMC with experimental reference data for an axisymmetric transonic bump. We provide detailed descriptions, in particular, of the calibration step and summary statistics used during the parameter estimation. Conclusions are provided at the end.

\section{Approximate Bayesian computation with Markov chain Monte Carlo sampling\label{sec:ABC}}
The task of turbulence model parameter estimation can be expressed as an inverse statistical problem where we must determine model parameters $\bm{c}$ that satisfy $\mathcal{F}(\bm{c})=\mathcal{D}$, where $\mathcal{F}$ is the model and $\mathcal{D}$ represents reference data from experiments or higher fidelity simulations. From a statistical perspective, this inverse problem can be solved by calculating the posterior distribution of $\bm{c}$ given $\mathcal{D}$ according to Bayes' theorem, which is expressed generically as
    \begin{equation}
	\label{eq:bayes_theorem}
	P(\bm{c}\,|\,\mathcal{D}) = \dfrac{L(\mathcal{D}\,|\,\bm{c})\pi(\bm{c})}{\int_{\bm{c}}L(\mathcal{D}\,|\,  \bm{c})\pi(\bm{c})d\bm{c}}\,.
\end{equation}
In the above theorem, $P(\bm{c}\,|\,\mathcal{D})$ is the posterior distribution, $L(\mathcal{D}\,|\, \bm{c})$ is the likelihood function, $\pi(\bm{c})$ is the prior distribution of model parameters, and ${\int_{\bm{c}}L(\mathcal{D}\,|\,  \bm{c})\pi(\bm{c})d\bm{c}}$ is a normalizing factor. A benefit of the Bayesian statistical approach is that $P(\bm{c}\,|\,\mathcal{D})$ naturally provides uncertainties associated with each estimated parameter, in contrast to other inversion techniques that provide only single-point estimates for unknown parameters. 

In order to obtain $P(\bm{c}\,|\,\mathcal{D})$, we must, in general, calculate each of the terms on the right hand side of Eq.~\eqref{eq:bayes_theorem}. The prior distribution $\pi(\bm{c})$ in Eq.~\eqref{eq:bayes_theorem} is typically formed using our knowledge of unknown parameters, and is often given (as in the current tests) by a uniform distribution with bounds that are sufficiently broad to contain the true parameter values.  

The primary challenge in solving Eq.~\eqref{eq:bayes_theorem} is then the calculation the likelihood function $L(\mathcal{D}\,|\,\bm{c})$, which gives the probability of obtaining the reference data for different choices of model parameters. Typically, this function is analytically intractable and computationally expensive to compute exactly, and so approximations (e.g., \cite{cheung_bayesian_2011,oliver_bayesian_2011}) or surrogate modeling (e.g., \cite{ray_bayesian_2016, ray_learning_2018, edeling_bayesian_2014}) are often used instead. Likelihood-free methods, by contrast, provide an alternative, and potentially highly flexible, method to obtain $P(\bm{c}\,|\,\mathcal{D})$, and here we use one such method---namely, ABC---to directly approximate the posterior without requiring \emph{a priori} knowledge of the likelihood function.

The ABC method was introduced, and first widely applied, in population genetics~\cite{beaumont2002,marjoram_markov_2003,wegmann_efficient_2009}, and was subsequently implemented in a wide range of other scientific areas~\cite{wawrzynczak2018,cameron2012, picchini2014,luciani2009epidemiological,zheng2017, beaumont2010} (detailed reviews are provided by~\citet{csillery2010}, \citet{marin_approximate_2012}, \citet{lintusaari2017fundamentals}, \citet{sisson2018handbook}, and \citet{beaumont2019}). More recently, ABC has been employed in engineering contexts for the estimation of rate coefficients in chemical kinetic models \cite{khalil2018probabilistic}, for the estimation of boundary conditions in complex thermal-fluid flows \cite{christopher2018}, and for determining unknown model parameters in autonomic \cite{doronina2018} and nonlinear \cite{doronina2020} subgrid-scale closure models for LES. \citet{doronina2019} were the first to take advantage of ABC-MCMC for discovering model parameter values and uncertainties in a multi-parameter RANS closure, using relatively simple homogeneous test cases as reference data for calibration of a nonequilibrium turbulence model \cite{hamlington_reynolds_2008,hamlington_frequency_2009,hamlington2014} consisting of three coupled ordinary differential equations (ODEs).

\subsection{Classic ABC rejection algorithm}
In general terms, the ABC algorithm samples model parameters $\bm{c}$ from the prior distribution $\pi(\bm{c})$ and compares model outcomes, $\mathcal{D}'=\mathcal{F}(\bm{c})$, with the reference data, $\mathcal{D}$. In order to reduce the computational expense of the approach, ABC makes use of summary statistics, $\mathcal{S}(\mathcal{D})$ and $\mathcal{S}'=\mathcal{S}(\mathcal{D}')$, to compare the model and reference data. Summary statistics can be a mean value, a standard deviation, a probability density function (pdf) or a subsample of the full data. The summary statistic comparison is performed using a statistical distance function, $d(\mathcal{S},\mathcal{S}')$, such as the Kullback-Leibler divergence, the Hellinger distance, or simply the mean-square error.

To obtain an estimate of the posterior from the comparison of the summary statistics, if the modeled and reference summary statistics are similar to within some specified tolerance $\epsilon$, then the sampled parameter values are considered to be sampled from the posterior distribution. Reducing the tolerance $\epsilon$ improves the approximation of the posterior, but also significantly increases the number of samples required by the algorithm. It should be noted that ABC is based on two main assumptions: (\emph{i}) that the summary statistic is sufficient, which means that the posterior distribution using the full data and the posterior distribution using summary statistics are the same, such that $P(\bm{c}\,|\,\mathcal{S}) = P(\bm{c}\,|\,\mathcal{D})$; (\emph{ii}) if the threshold $\epsilon$ is sufficiently small (i.e., as $\epsilon \rightarrow 0$), then ABC produces the exact posterior distribution, $P(\bm{c}\,|\,\mathcal{S}) = \lim_{\varepsilon\rightarrow 0} P\left[\bm{c}\,|\, d(\mathcal{S}^{\prime}, \mathcal{S}) \le \varepsilon\right]$. 

The algorithm described above is called the rejection ABC algorithm (or classic ABC) and is provided in Appendix \ref{appendix} and shown schematically in Fig.~\ref{fig:algorithm}. 

\subsection{Improved ABC algorithm with Markov-chain Monte Carlo sampling}
The classic ABC rejection algorithm described in the previous section can be computationally expensive, since the number of sampled parameters grows exponentially with the number of unknown parameters, $N$, in the model. The computational cost of ABC can be significantly improved by using the MCMC method for sampling model parameters. The ABC-MCMC method (or MCMC without likelihood), introduced by \citet{marjoram_markov_2003}, samples the next parameter from the proposal distribution $q(\bm{c}_i\rightarrow\bm{c}')$. Then, if the statistical distance $d$ is within the tolerance $\epsilon$, the proposed parameters are accepted with probability 
\begin{equation}
h = \min\left[1, \frac{\pi(\bm{c}')q(\bm{c}_i\rightarrow \bm{c}')} {\pi(\bm{c}_i)q(\mathbf{c'}\rightarrow \bm{c}_i)}\right]\,.
\end{equation}
In practice, this is implemented by randomly selecting a number between 0 and 1, which bound $h$, and then checking whether this random number is less than $h$. 

The choice of proposal distribution $q$ is crucial for the convergence rate of the ABC-MCMC algorithm. A variable proposal with an adapting size and spatial orientation provides faster convergence, and here we follow the adaptive proposal procedure outlined by \citet{haario_adaptive_2001}, where $q(\bm{c}_i\rightarrow \bm{c}') = q(\bm{c}'| \bm{c}_i,\mathcal{C}_i)$ is the Gaussian proposal with the current parameter $\bm{c}_i$ as the mean value and adaptive covariance $\mathcal{C}_i$. The covariance $\mathcal{C}_i$ is updated during the process using all previous steps of the chain.

\begin{figure}[!t]
	\centering
    \includegraphics[width=\linewidth]{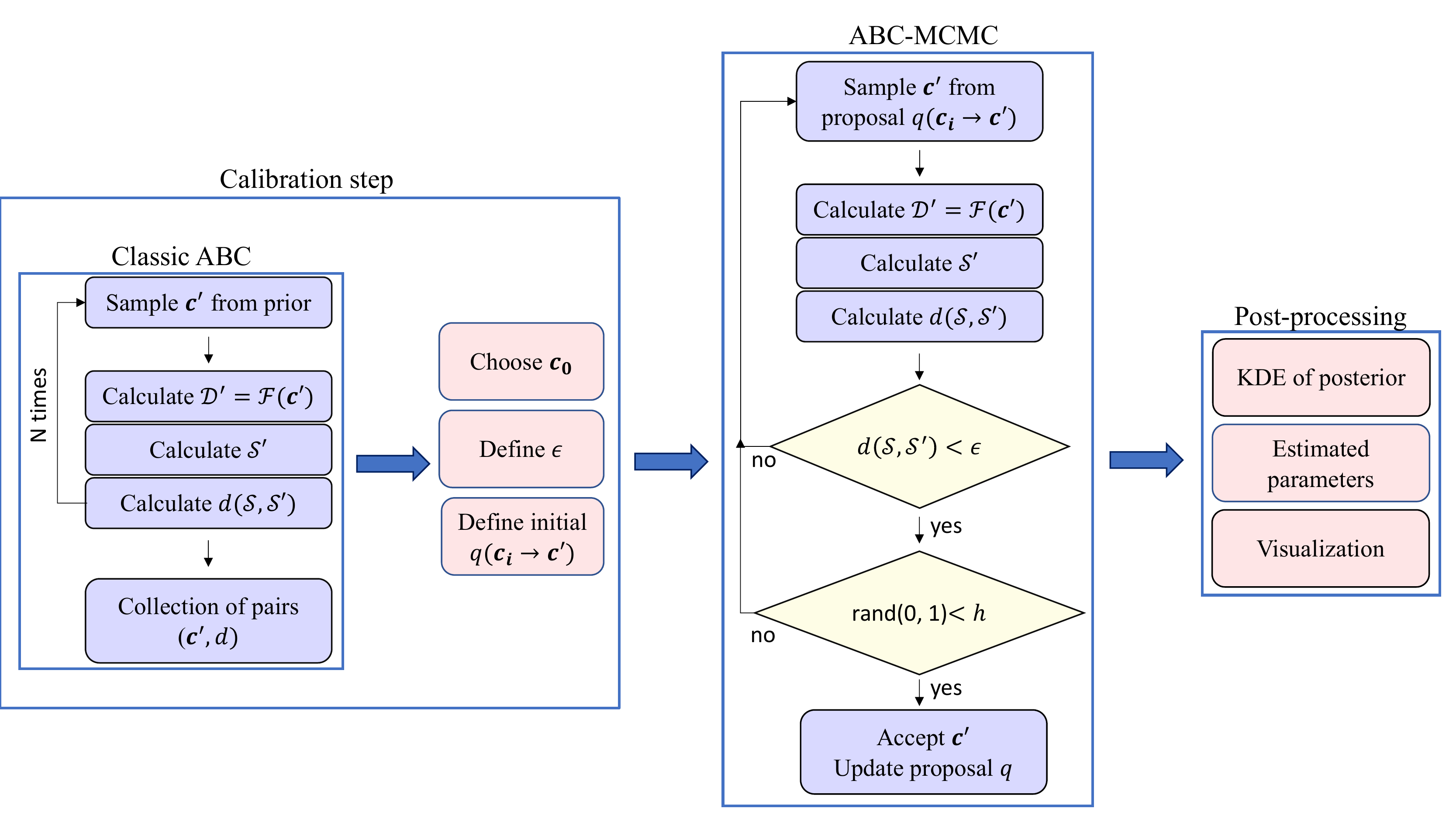}
	\caption{Schematic representation of the ABC-IMCMC algorithm (also provided in Appendix \ref{appendix}), showing the calibration step, core MCMC approach, and post-processing steps.}
	\label{fig:algorithm}
\end{figure}

We also adopt a calibration step before performing the ABC-MCMC algorithm, introduced by \citet{wegmann_efficient_2009}.  This adjusts the range of parameters and starting point for the chains, and defines initial parameters for adaptive proposal $q$. The final ABC-MCMC algorithm with calibration step and adaptive proposal is termed the improved ABC-MCMC algorithm, or ABC-IMCMC, and is provided in Appendix \ref{appendix} and schematically summarized in Fig.~\ref{fig:algorithm}. Additional details of the ABC-IMCMC algorithm can also be found in \citet{doronina2020}.

It should be noted that any particular implementation of ABC-IMCMC requires the specification of the model, reference data, summary statistics, statistical distance function, and distance threshold. Specific choices for each of these quantities are outlined in the demonstrations presented in Sections \ref{sec:nonequil_result} and \ref{sec:overflow}, and the generality of the resulting model parameters can be increased through the inclusion of more, and increasingly different, reference data. Eventually, for sufficiently broad reference data sources, the accuracy of the resulting model will be limited by the model form itself, rather than by any limitations intrinsic to the ABC-IMCMC method. 

\subsection{Post-processing of ABC results}
As a result of either the classic ABC or ABC-IMCMC algorithms, we obtain a list of ``accepted'' parameter values (i.e., parameter values that are sampled from the approximate posterior distribution). To estimate this approximated joint posterior, we can build an $N$-dimensional histogram, but the different choices of bin size and location can lead to distributions that have qualitatively different features, particularly when the number of counts in the histogram is small (as can be the case for small values of the threshold $\epsilon$). 

To avoid this dependence on bin size and location in the $N$-dimensional posteriors, we use Gaussian Kernel Density Estimation (KDE) to represent each parameter sample with a Gaussian, and thereby obtain a continuous replacement for the discrete histogram. The standard deviation of the Gaussian kernel (i.e., the kernel bandwidth) determines the smoothness of the estimated distribution; here we use Scott's rule~\cite{scott2015multivariate, silverman1986density} to define the bandwidth in Gaussian KDE. To apply Gaussian KDE to the $N$-dimensional posteriors, we use the \texttt{FFTKDE} function from the \texttt{KDEpy} python package.\footnote{\url{https://github.com/tommyod/KDEpy}}

To visualize the $N$-dimensional posterior, we plot one- and two-dimensional marginalized pdfs produced by Gaussian KDE. Single estimates of unknown parameters are then taken from the $N$-dimensional posterior as the maximum \textit{a posteriori} probability (MAP) values. Bayesian confidence intervals for the MAP estimates can be directly computed from the $N$-dimensional posterior, and we can also propagate the posterior distribution to make uncertainty predictions for the quantity of interest.

\section{Demonstration of ABC: Nonequilibrium RANS model \label{sec:nonequil_result}}
We first demonstrate the use of the classic ABC rejection algorithm (Algorithm~\ref{alg:abc-rej} in Appendix \ref{appendix}) to estimate parameters in a nonequilibrium RANS model for predictions of kinetic energy time series in periodically sheared homogeneous turbulence. Two different reference data sources are considered here. The first source is the nonequilibrium RANS model itself, thereby testing the ability of the ABC algorithm to recover known parameter values (i.e., a verification test). This verification test demonstrates how well we can recover nominal parameter values when the reference data is free from measurement error and can be exactly recreated by the model; a similar test was performed for the estimation of boundary conditions in LES by \citet{christopher2018}. The second reference data source is provided by the DNS study of \citet{yu_direct_2006}, showing the ability of ABC to estimate parameters based on higher-fidelity computational reference data (i.e., a validation test). In this case, the model and reference data are independent, and this test demonstrates the ability of ABC to recover unknown parameter values when the model is not expected to perfectly match the reference data.

It should be noted that, because the nonequlibrium RANS model can be implemented in the present homogeneous tests as simply a one-dimensional (in time) integration of three coupled ODEs, the additional acceleration provided by the full ABC-IMCMC method is not required. The more efficient ABC-IMCMC method is demonstrated in Section \ref{sec:overflow}, where it is used to estimate the joint posterior distribution of parameter values in the Menter SST RANS model for an inhomogeneous axisymmetric transonic bump.

Finally, before continuing, we note that the parameter values estimated here should not be taken as universal for the nonequilibrium RANS model. It is likely that other sources of reference data, and even other choices of summary statistic, will yield different parameter estimates. However, in Section \ref{sec:overflow} we show how ABC-IMCMC can be used with multiple summary statistics to increase the generality of the resulting parameter values, and multiple different physical configurations and flows can be incorporated in a single ABC parameter estimation in future work.

\subsection{Nonequilibrium RANS model\label{subsec:noneq_model}}
The unclosed Reynolds stress appearing in the ensemble averaged RANS equations can be written in terms of isotropic and anisotropic (or deviatoric) parts as $\overline{u'_i u'_j} = (2/3)k\delta_{ij}+k a_{ij}$ where $k$ is the turbulence kinetic energy and $a_{ij}$ is the Reynolds stress anisotropy tensor. Because the governing equation for $k$ includes relatively few unclosed terms, the primary challenge in RANS modeling is to represent the tensor $a_{ij}$, which is unclosed. A number of models have been proposed for this tensor, including various nonlinear eddy viscosity models \cite{girimaji1996,Wallin2000,gatski2000}, but by far the most widely used models continue to be equilibrium models where it is assumed that $a_{ij} \propto \overline{S}_{ij}$. Such models include the classical $k$-$\varepsilon$ and $k$-$\omega$ models, as well as their many variants (see \citet{speziale1998} for a review).

However, it is now widely understood that equilibrium models become inaccurate in complex flows, and in particular those flows with rapid spatial and temporal variations in mean flow properties \cite{hamlington_reynolds_2008}. Such ``nonequilibrium'' turbulent flows arise in many practical applications, including supersonic and hypersonic vehicles where interactions occur between shock waves and turbulent boundary layers, as well as in internal combustion engines where pistons rapidly strain the flow. Nonequilibrium effects introduced in such flows can be significant, requiring new modeling approaches. 

The nonequilibrium turbulence model considered in this demonstration of ABC was identified by \citet{hamlington2014} as a nearly ideal model, in terms of both accuracy and computational simplicity, for capturing the effects of rapid straining in piston-driven and rapidly-compressed flows. This model is expressed for an incompressible flow as
\begin{equation}
    \label{eq:noneq}
    \frac{\partial a_{ij}}{\partial t} + \overline{u}_k \frac{\partial a_{ij}}{\partial x_k}= - \alpha_1 \frac{\varepsilon}{k} a_{ij} + \alpha_2 \overline{S}_{ij}\,,
\end{equation}
where $\overline{u}_i$ is the mean velocity, $\overline{S}_{ij} = (1/2)(\partial \overline{u}_i/\partial x_j + \partial \overline{u}_j/\partial x_i)$ is the mean strain rate tensor, $\varepsilon$ is the kinetic energy dissipation rate, and the coefficients $\alpha_1$ and $\alpha_2$ are given by
\begin{equation}
\label{eq:alpha}
\alpha_1 = \frac{P_k}{\varepsilon}-1+C_1\,,\quad \alpha_2 = C_2 - \frac{4}{3}\,,
\end{equation}
where $P_k= -k a_{ij} \overline{S}_{ij}$ is the kinetic energy production rate. Parameters $C_1$ and $C_2$ in Eq.~\eqref{eq:alpha} have traditionally been tied to the choice of pressure-strain correlation model \cite{launder1975,speziale1991a}, but here they are treated as unknowns determined by ABC. 

It should be noted that the model in Eq.~\eqref{eq:noneq} has a rigorous basis in the exact anisotropy transport equation and accounts for both the return to isotropy of unstrained turbulence [i.e., the first term on the right hand side of Eq.~\eqref{eq:noneq}] and the generation of anisotropy in strained turbulence [i.e., the second term on the right hand side of Eq.~\eqref{eq:noneq}]. This model is also the basis for the quasi-analytical nonequilibrium anisotropy model outlined in \cite{hamlington_reynolds_2008, hamlington_frequency_2009}. 

The test case considered here is homogeneous, which permits substantial simplifications to the RANS and anisotropy equations. In particular, since spatial derivatives of turbulent fluctuating variables are identically zero in homogeneous turbulence, the Reynolds stresses have no effect on the evolution of $\overline{u}_i$, and the spatial derivative of $a_{ij}$ on the left hand side of Eq.\ \eqref{eq:noneq} is identically zero. As a result, for homogeneous flows where $\overline{S}_{ij}$ varies in time only, the evolution of $a_{ij}$ is given by
\begin{equation}
    \label{eq:noneq1}
    \frac{d a_{ij}}{d t} = \left(\frac{k}{\varepsilon}a_{ij}\overline{S}_{ij}+1-C_1\right) \frac{\varepsilon}{k} a_{ij} + \left(C_2 - \frac{4}{3}\right) \overline{S}_{ij}\,,
\end{equation}
where $a_{ij}=a_{ij}(t)$ is a function of time only and Eq.~\eqref{eq:alpha} has been used to replace the $\alpha_i$ coefficients appearing in Eq.~\eqref{eq:noneq}. Similarly, in homogeneous turbulent flows, $k=k(t)$ and $\varepsilon=\varepsilon(t)$ also depend only on time and their dynamics are represented here using the standard equations employed in classical $k$-$\varepsilon$ models, namely
\begin{align}
\label{eq:k}
\frac{dk}{dt} &= -ka_{ij}\overline{S}_{ij} - \varepsilon\,, \\
\label{eq:eps}
\frac{d\varepsilon}{dt} &= -C_{\varepsilon1}\varepsilon a_{ij}\overline{S}_{ij} - C_{\varepsilon2}\frac{\varepsilon^2}{k}\,,
\end{align}
where $C_{\varepsilon1}$ and $C_{\varepsilon2}$ are additional unknown parameters that will be estimated using the ABC procedure.

The system of nonlinear and coupled ODEs represented by Eqs.~\eqref{eq:noneq1}--\eqref{eq:eps} constitutes the nonequilibrium turbulence anisotropy closure examined in this demonstration of ABC, and depends on the unknown model parameters $\bm{c} = (C_1, C_2, C_{\varepsilon1}, C_{\varepsilon2})$. For a given choice of $\overline{S}_{ij}$ and initial conditions for $a_{ij}$, $k$, and $\varepsilon$, Eqs.~\eqref{eq:noneq1}--\eqref{eq:eps} can be straightforwardly integrated as a system of coupled nonlinear ODEs.

\subsection{Periodically sheared homogeneous turbulence\label{subsec:homogeneous_cases}}
The nonequilibrium anisotropy closure described in Section \ref{subsec:noneq_model} is applied here to predict the evolution of the turbulence kinetic energy in periodically sheared homogeneous turbulence \cite{yu_direct_2006}. For this case, the turbulence is assumed to be initially isotropic and unstrained such that $a_{ij}=0$ and $\overline{S}_{ij}=0$ for $t<0$.  For $t\geq 0$, the turbulence is then subjected to a periodic mean strain rate tensor $\overline{S}_{ij}$ with a shearing frequency $\omega$, given by
\begin{equation}
\overline{S}_{ij} =\frac{S}{2} \begin{bmatrix}
0   &  \sin(\omega t)  &0 \\
\sin(\omega t) &0   & 0 \\
0   &0      &0
\end{bmatrix}.
\end{equation}
At time $t=0$, it is assumed that $k=k_0$ and $\varepsilon=\varepsilon_0$, and the initialization of the case is completed by defining $S k_0/\varepsilon_0 = 3.3$ and $\omega/S = 0.5$. The specific case of $\omega/S = 0.5$ is chosen for the present demonstration since this shearing frequency is close to the frequency at which the system transitions to a fully nonequilibrium state, as described in \citet{hamlington_frequency_2009}, and is thus anticipated to provide a challenging inverse modeling problem for ABC. Using this strain rate time series and initial conditions, the model data, $\mathcal{D}'=\mathcal{F}(\bm{c})$, for the ABC rejection algorithm are provided by integrating the system of coupled ODEs in Eqs.~\eqref{eq:noneq1}--\eqref{eq:eps} using the \texttt{ODEINT} function from \texttt{scipy.integrate}~\cite{2020SciPy-NMeth}.

\subsection{Setup of ABC}
In the following, we verify and validate the ABC rejection algorithm (Algorithm~\ref{alg:abc-rej} in Appendix A) by estimating model parameters for the nonequilibrium anisotropy closure represented by Eqs.~\eqref{eq:noneq1}--\eqref{eq:eps}. Reference data for this case are taken from two sources: (\emph{i}) an integration of the nonequilibrium model using the nominal parameter values $\bm{c} = (1.5, 0.8, 1.44, 1.83)$ \cite{hamlington_frequency_2009}, and (\emph{ii}) the study by \citet{yu_direct_2006}, where DNS was performed for a series of different values of $\omega/S$. 

The reference and modeled summary statistics, $\mathcal{S}$ and $\mathcal{S}'$, respectively, for both the verification and validation cases are given by the specific values of the turbulence kinetic energy, $k_i=k(t_i)$, at the times $t=t_i$ when the DNS reference data from \citet{yu_direct_2006} are provided. This summary statistic thus represents a subset of all data $\mathcal{D}'$ produced by the model for any choice of $\bm{c}$. 

The distance function is defined simply as the 2-norm of the difference between the reference and modeled summary statistics, namely
\begin{equation}
    d(\mathcal{S}, \mathcal{S}') =  \left[\sum_{i}\left(k^{\prime }_{i} - k_{i}\right)^2\right]^{1/2}\,.
\end{equation}
Here $k'_i$ and  $k_i$  indicate the turbulence kinetic energy from the model and reference data, respectively.

To construct the posterior distribution, we uniformly sample $N = 12,960,000$ parameters $\bm{c}$ (given by a uniform grid with 60 samples per dimension) and calculate the distance function between the modeled and reference summary statistics for each of the sampled parameters. Saving all calculated distances, we choose the desired acceptance rate by changing $\epsilon$. If $\epsilon$ is sufficiently small, the approximated posterior resulting from the ABC algorithm recovers the exact posterior distribution. In practice, however, an $\epsilon$ that is too small leads to too few sampled parameters passing the acceptance threshold, $d(\mathcal{S}, \mathcal{S}')\le \epsilon$, resulting in a poorly converged posterior. Conversely, relaxing the acceptance criterion too much (i.e., using a larger $\epsilon$) can lead to the final posterior distribution being biased towards the prior. Thus, the hyperparameter $\epsilon$ should be reasonably small but should also take into account the computational cost of every model evaluation. For both the verification and validation tests presented here, we consider a range of $\epsilon$ values to show the convergence of the posterior distribution. 

\begin{figure}[t!]
    \includegraphics{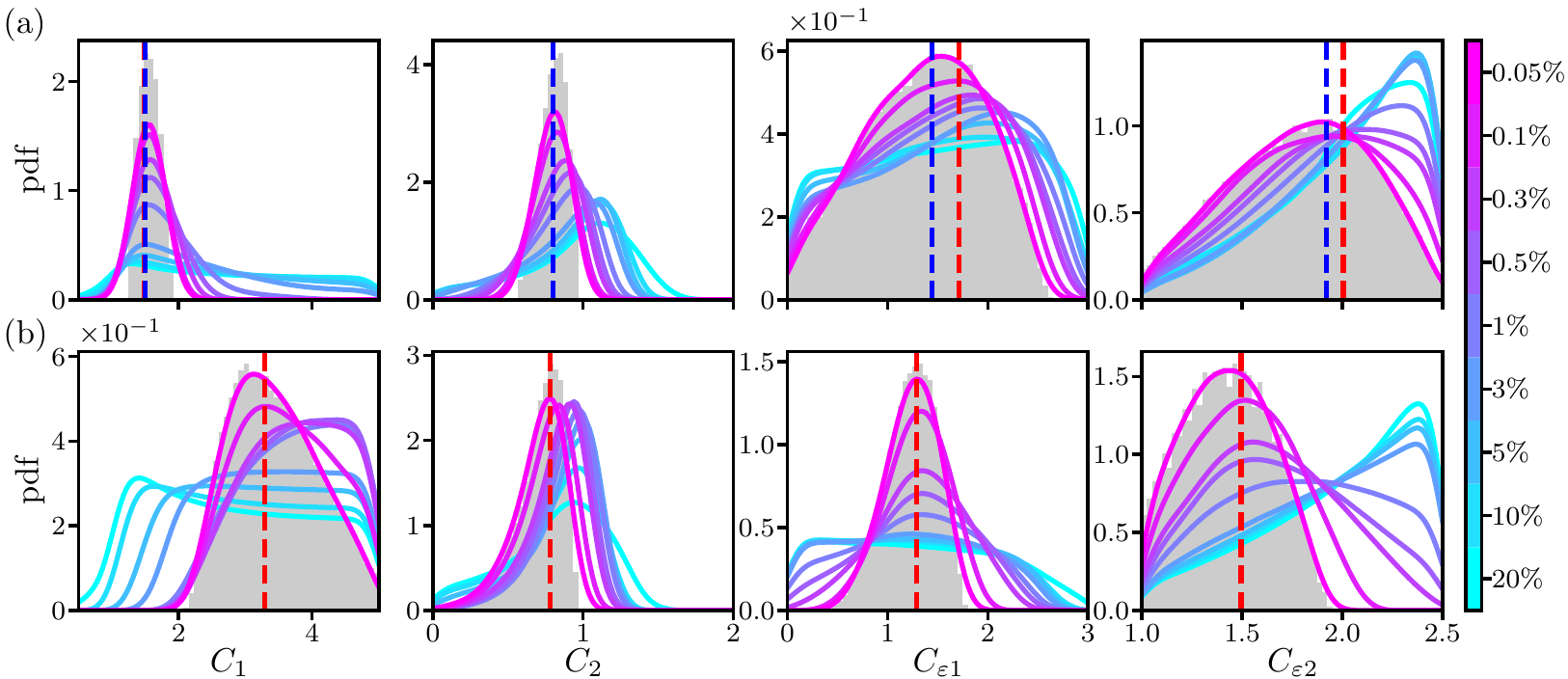}%
    \caption{Marginal posteriors for parameter estimation in the nonequilibrium RANS model for (a) the verification case using reference data from the nonequilibrium RANS model and (b) the validation case using reference data from the DNS study by \citet{yu_direct_2006}. Line colors correspond to the percentage of accepted simulations, as determined by the rejection threshold $\epsilon$. Raw marginals for a 0.05\% acceptance rate (i.e., before Gaussian kernel density estimation) are shown by gray bars. Vertical blue dashed lines show the nominal parameter values and vertical red dashed lines show the estimated MAP values.} 
    \label{fig:rans_ode_marginal}
\end{figure}

\subsection{Results and discussion}
Figure~\ref{fig:rans_ode_marginal} shows marginal posterior distributions for $C_1$, $C_2$, $C_{\varepsilon1}$, and $C_{\varepsilon2}$ for both the verification and validation cases. All distributions are constructed using Gaussian KDE with bandwidths defined by Scott's rule. Marginal posteriors for different values of $\epsilon$ are shown in Fig.~\ref{fig:rans_ode_marginal}, with corresponding acceptance rates changing from 20\% to 0.05\% as $\epsilon$ decreases. The resulting posteriors are shown to converge as the acceptance rate decreases.

Estimates of the unknown parameters are obtained as the MAP values of the four-dimensional posterior distribution with a 0.05\% acceptance rate. For the verification test shown in Fig.~\ref{fig:rans_ode_marginal}(a), the MAP values agree closely with the nominal values (i.e., the values used to generate the reference data) for $C_1$ and $C_2$, and the ratio of $C_{\varepsilon2}$ to $C_{\varepsilon1}$ is the same for both the model and reference data (more discussion of this point is provided below). This demonstrates the ability of the ABC rejection algorithm to recover expected parameter values in the verification test. 

\begin{figure}[t!]
    \centering\includegraphics{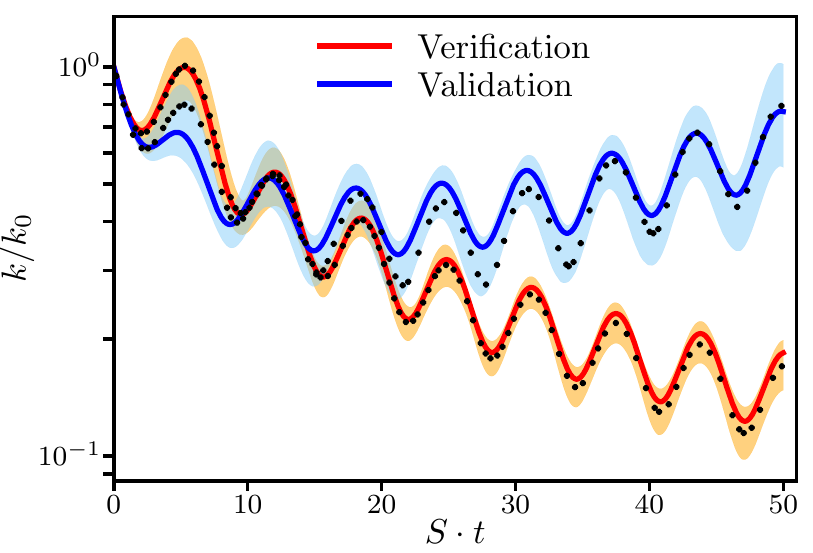}
    \caption{Turbulence kinetic energy $k(t)/k_0$ for the nonequilibrium model with estimated parameters (solid lines) and reference data (points) for the periodically sheared turbulence case. The shaded areas indicate the 99\% confidence intervals. The verification case (red line) uses reference data from the nonequilibrium RANS model, while the validation case uses reference data from the DNS study by \citet{yu_direct_2006}.}\label{fig:rans_ode_MAP}
\end{figure}

Results for the validation test in Fig.~\ref{fig:rans_ode_marginal}(b), where the reference data is from the DNS, again show a convergence of the posteriors as $\epsilon$ decreases. The MAP estimates of this case for $C_1$ and $C_{\varepsilon2}$ are larger and smaller, respectively, than the nominal values. These variations are responsible for weakening the relaxation towards isotropy (in the case of $C_1$) and reducing the dissipation of $k$ (in the case of $C_{\varepsilon2}$); this will be shown in Fig.~\ref{fig:rans_ode_MAP} to result in larger magnitudes of $k$ as compared to results using the nominal parameter values. Figure \ref{fig:rans_ode_marginal} shows that $C_2$ and $C_{\varepsilon1}$ remain close to the nominal values. We note that the nominal value of $C_2$ has a fundamental physical justification based on consideration of rapidly strained turbulence \cite{crow_1968,hamlington2009a}, and so the similarity of the estimated value to the nominal value when using ABC with DNS reference data is perhaps not unexpected (although it is also not guaranteed). 

The accuracy of the parameters estimated using ABC is indicated in Fig.~\ref{fig:rans_ode_MAP}, which shows the evolution of the turbulence kinetic energy $k(t)/k_0$ for the verification and validation cases. Results for the verification case are in very good agreement with the reference data, again indicating that ABC can accurately recover parameter values used to generate the reference data. The results for the validation case do not match quite as closely with the DNS reference data, but the overall agreement is reasonable and the initial decrease, followed by a subsequent increase, of $k$ is captured by the model; this behavior is due, in part, to the larger value of $C_1$ and smaller value of $C_{\varepsilon2}$ estimated by ABC, as compared to the nominal values. Most importantly, the parameter values estimated from ABC recover the DNS data much more accurately than do the nominal values for the model. We can propagate the uncertainty through the model to estimate uncertainty intervals on the output quantity of interest. Figure~\ref{fig:rans_ode_MAP} shows the 99\% confidence interval for the modeled kinetic energy, and this confidence interval encompasses the DNS reference data used in the validation case.

\begin{figure}[t!]
	\centering\includegraphics{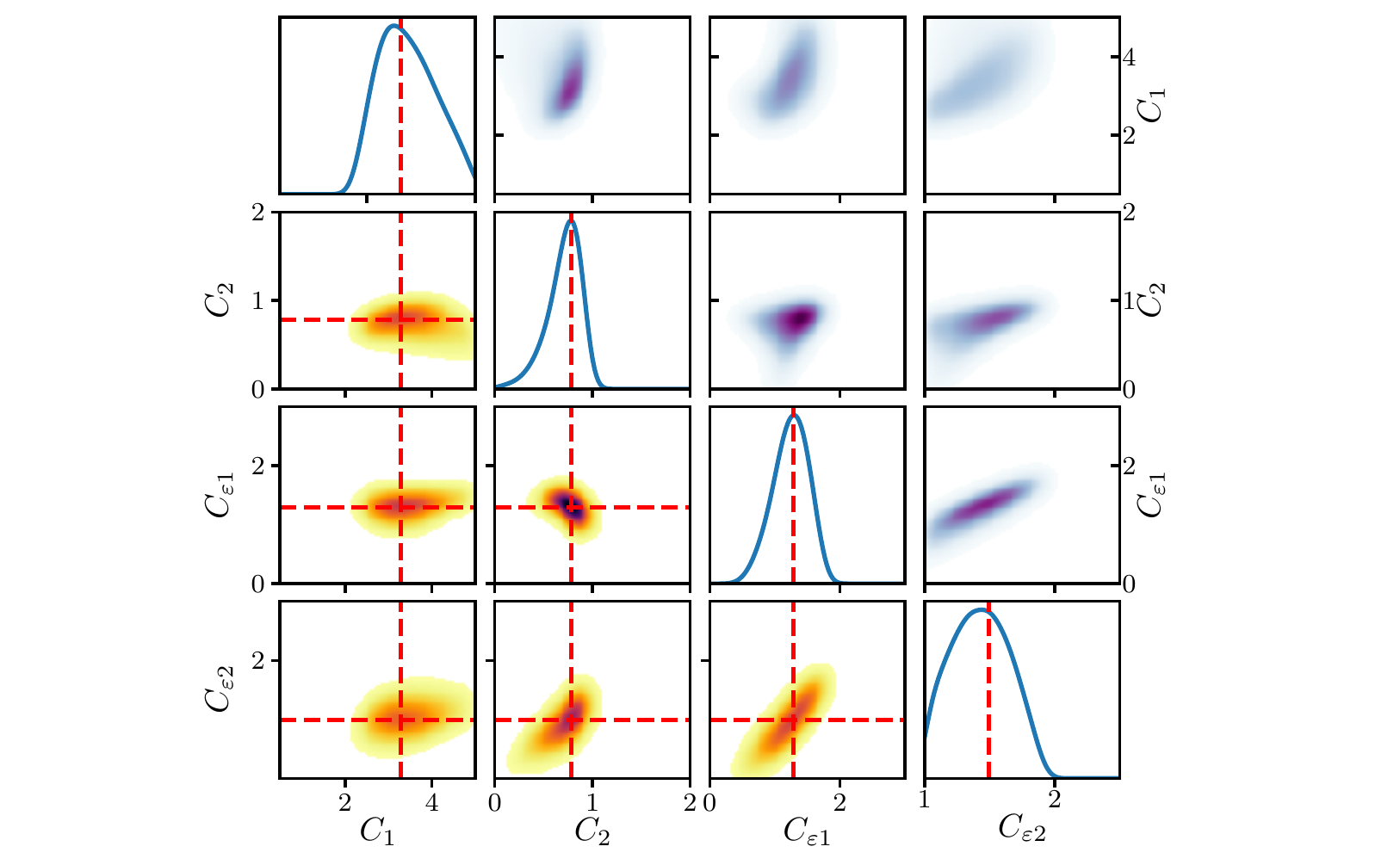}
	\caption{Posterior distributions of accepted parameters for the nonequilibrium RANS model for the validation case using reference data from the DNS study by \citet{yu_direct_2006}. Diagonal subplots show one-dimensional marginal pdfs, upper-diagonal subplots show 2D marginalized pdfs, and lower-diagonal subplots show conditional pdfs taken at the MAP values. The red dashed lines represent estimated parameters taken from the MAP values of the joint posterior.}
	\label{fig:rans_ode_2d_marginals}
\end{figure}

In order to facilitate visualization of the four-dimensional posteriors resulting from the ABC procedure for the nonequilibrium model, marginal pdfs for the validation case are shown on the diagonal subplots in Fig.~\ref{fig:rans_ode_2d_marginals} for each of $C_1$, $C_2$, $C_{\varepsilon1}$, and $C_{\varepsilon2}$; posteriors for the verification case are similar to those from the validation case and are consequently not shown here. Two dimensional marginalized joint pdfs are shown on the upper-diagonal subplots and conditional pdfs taken at the MAP values are shown on the lower-diagonal subplots of Fig.~\ref{fig:rans_ode_2d_marginals}. These two-dimensional marginalized joint pdfs allow us to identify parameter correlations that may not be obvious from the marginalized posteriors in Fig.~\ref{fig:rans_ode_marginal}. 

In particular, Fig.~\ref{fig:rans_ode_2d_marginals} shows that $C_{\varepsilon1}$ and $C_{\varepsilon2}$ are closely correlated for the validation case, and the same is also true for the verification case. This explains the shift in the MAP values of these two parameters indicated for the verification case in Fig.~\ref{fig:rans_ode_marginal}(a). That is, even though both $C_{\varepsilon1}$ and $C_{\varepsilon2}$ are shown to be larger than their nominal values, their ratio is nearly identical to the ratio of the nominal values. This indicates that only the ratio of these two parameters is important in the present tests. The importance of this ratio is consistent with the analysis of the periodically sheared case by \citet{hamlington_frequency_2009}, where it was shown that $(C_{\varepsilon2}-1)/(C_{\varepsilon1}-1)$ was a parameter of fundamental importance in determining evolution of the production to dissipation rate $P_k/\varepsilon$ and the transition to the saturated nonequilibrium state for sufficiently large shearing frequencies.

\section{Demonstration of ABC-IMCMC: Menter SST RANS model\label{sec:overflow}}
The demonstration in Section \ref{sec:nonequil_result} shows the effectiveness of ABC for model parameter estimation in a relatively simple homogeneous flow. In the vast majority of cases, however, we would instead like to use ABC with reference data from more complex inhomogeneous flows, for which each model evaluation during the ABC procedure is more expensive. This additional computational cost motivates the use of the more efficient ABC-IMCMC algorithm, summarized in Fig.~\ref{fig:algorithm} and Appendix A.

In the following, we use ABC-IMCMC to estimate parameters in the Menter SST RANS model, given experimental reference data for an axisymmetric transonic bump. We simultaneously use summary statistics based on the mean velocity, shear stress, pressure, and separation and reattachment points, demonstrating the flexibility of the ABC-IMCMC approach. Due to the computational cost of this test case, we perform an initial calibration step to determine which of the unknown SST model parameters to include in the estimation process, in addition to initializing the Markov chains. It will be seen that this initial step allows us to reduce the number of unknown parameters estimated in the model, thereby substantially reducing the computational cost. Final parameter estimates from ABC-IMCMC and comparisons with experimental data are provided at the end of this section.

Before continuing, we once again emphasize that the demonstration in this section is intended to be illustrative of the full ABC-IMCMC approach for an inhomogeneous flow where the model data are provided by forward runs using a CFD code. The resulting estimated parameters should, therefore, not be taken as universal, although the generality of the model can be improved in the future through the simultaneous use of reference data from many different flows; the ABC-IMCMC procedure is sufficiently flexible to accommodate a range of data types and sources.

\subsection{Menter SST RANS model\label{sec:menter_sst}}
The classical Menter SST model is a two-equation eddy-viscosity RANS model introduced by~\citet{menter1994standard} in~\citeyear{menter1994standard}. It is based on the baseline model~\cite{menter1994standard}, which combines the Wilcox $k-\omega$ model~\cite{wilcox1988} in the near-wall region and a standard $k-\varepsilon$ model in the wake region of the boundary layer. This model blending was introduced to take advantage of the freestream independence of the $k-\varepsilon$ model in the outer part of the boundary layer, combined with the superior behavior of the $k-\omega$ model in the logarithmic part of the boundary layer in compressible flows and equilibrium flows with adverse pressure gradients. For convenience, the $k-\varepsilon$ model is transformed into the $k-\omega$ formulation, which differs from the original $k-\omega$ model by a cross-diffusion term in the $\omega$ equation, and also has different coefficients. Using the same transport equations, the Menter SST model incorporates Bradshaw's hypothesis that the principal turbulent shear-stress is proportional to the turbulent kinetic energy $k$.

In this study, we follow the nomenclature of the  NASA Langley Turbulence Modeling Resource,\footnote{\url{https://turbmodels.larc.nasa.gov/sst.html}} and the complete formulation and explanation of the Menter SST model can be found in \citet{menter1994standard}. Here we repeat the key components of the model in order to identify the parameters that will be estimated in the present ABC-IMCMC approach. The transport equations for $k$ and $\omega$ are 
\begin{align}\label{eq:model_k}
\frac{\partial(\rho k)}{\partial t} + \frac{\partial(\rho \overline{u}_j k)}{\partial x_j} &= P_k - \beta^*\rho \omega k + \frac{\partial}{\partial x_j}\left[(\mu + \sigma_k\mu_t)\frac{\partial k}{\partial x_j}\right]\, , 
\\\label{eq:model_omega}
\frac{\partial(\rho \omega)}{\partial t} +  \frac{\partial(\rho \overline{u}_j \omega)}{\partial x_j} &= \frac{\gamma}{\nu_t}P_k - \beta\rho\omega^2 + \frac{\partial}{\partial x_j}\left[(\mu + \sigma_{\omega}\mu_t)\frac{\partial \omega}{\partial x_j}\right] +2(1-F_1)\rho\sigma_{\omega2}\frac{1}{\omega}\frac{\partial k}{\partial x_j}\frac{\partial \omega}{\partial x_j}\, ,
\end{align}
where $\rho$ is the density, $\nu_t = \mu_t/\rho$ is the turbulent kinematic viscosity and $F_1$ is the blending function, such that $F_1 = 1$ in the near-wall region, activating the original $k-\omega$ model, and $F_1=0$ away from the surface, activating the transformed $k-\varepsilon$ model. The $F_1$ function is defined as
\begin{equation}
F_1 = \tanh(\mathrm{arg}_1^4)\,,\quad\mathrm{arg}_1 = \min\left[\max\left(\frac{\sqrt{k}}{\beta^*\omega y}, \frac{500\nu}{y^2\omega}\right), \frac{4\rho\sigma_{\omega2}k}{CD_{k\omega}y^2}\right]\,,\quad
 CD_{k\omega} = \max{\left(\frac{2\rho\sigma_{\omega2}}{\omega}\frac{\partial k}{\partial x_j}\frac{\partial \omega}{\partial x_j}, 10^{-20}\right)}\, ,
\end{equation}
where $y$ is the distance from the wall and $\mathrm{arg}_1$ goes to zero far from walls because of the $y$ and $y^2$ factors in the denominators of all three terms. Thus, the cross-diffusion term with $(1-F_1)$ in Eq.~\eqref{eq:model_omega} disappears near the wall. The coefficients $\phi = (\gamma, \beta, \sigma_k, \sigma_{\omega})$ are different for each part of the blended model, with $\phi_1$ denoting values for the $k-\omega$ model and $\phi_2$ for the $k-\varepsilon$ model. The combined model in Eqs.~\eqref{eq:model_k} and \eqref{eq:model_omega} thus uses a single $\phi$ defined as
\begin{equation}
\phi = F_1\phi_1+(1-F_1)\phi_2\,.
\end{equation}
Closure coefficients and their nominal values are provided in Table~\ref{tab:SST_coef}. Other closure coefficients are defined using the values in Table~\ref{tab:SST_coef} as
\begin{equation}
\gamma_1 = \frac{\beta_1}{\beta^*} - \sigma_{\omega1}\frac{\kappa^2}{\sqrt{\beta^*}}\,, \qquad
\gamma_2 = \frac{\beta_2}{\beta^*} - \sigma_{\omega2}\frac{\kappa^2}{\sqrt{\beta^*}}\,.
\end{equation}
The turbulence kinetic energy production, $P_k$, in Eqs.~\eqref{eq:model_k} and \eqref{eq:model_omega} is defined as
\begin{equation}
P_k = -\rho\overline{u'_i u'_j}\frac{\partial \overline{u}_i}{\partial x_j}\, ,\quad
\rho\overline{u'_i u'_j} = \frac{2}{3}\rho k\delta_{ij}-\mu_t\left(\frac{\partial \overline{u}_i}{\partial x_j} + \frac{\partial \overline{u}_j}{\partial x_i} - \frac{2}{3}\frac{\partial \overline{u}_k}{\partial x_k}\delta_{ij}\right) \,,
\end{equation}
where $\mu_t$ is the turbulent (dynamic) eddy viscosity.

\begin{table}[t!]
	\centering
	\caption{Nominal values of Menter SST model coefficients.}
	\label{tab:SST_coef}
	\begin{tabular}{l|ccccccccc}
		\hline
		Coefficient   & $\sigma_{k1}$ & $\sigma_{k1}$ & $\sigma_{\omega1}$ & $\sigma_{\omega2}$ & $\beta_1$ & $\beta_2$ & $\beta^*$ & $\kappa$ & $a_1$ \\ \hline
		Nominal value & 0.85       & 1.0          & 0.5         & 0.856         & 0.075    & 0.828  & 0.09      & 0.41  & 0.31  \\
		\hline
	\end{tabular}
\end{table}

In the $k-\varepsilon$ and $k-\omega$ two-equation models, the principal shear-stress  $\tau\equiv-\rho\overline{u'v'}$ is usually computed as $\tau = \mu_t (\partial \overline{u}/ \partial y)$. However, based on Bradshaw's assumption, the shear stress in the boundary layer is proportional to the turbulent kinetic energy $k$ as $\tau = \rho a_1 k$, with $a_1$ being constant. Taking this into account, \citet{menter1994standard} defined the turbulent eddy viscosity as
\begin{equation}\label{eq:mu_sst}
\mu_t = \frac{\rho a_1k}{\max(a_1\omega, \Omega F_2)}\, ,
\end{equation}
where $\Omega = \sqrt{2R_{ij}R_{ij}}$ with $R_{ij} = 1/2\left( \partial \overline{u}_i/\partial x_j - \partial \overline{u}_j/\partial x_i\right)$. The second blending function, $F_2$, appearing in this definition for $\mu_t$ is given by
\begin{equation}
F_2 = \tanh(\mathrm{arg}_2^2)\, ,\quad \mathrm{arg}_2 = \max\left(\frac{\sqrt{k}}{\beta^*\omega y}, \frac{500\nu}{y^2\omega}\right)\, .
\end{equation}
In an adverse-pressure-gradient boundary layer, production of $k$ is larger than its dissipation (i.e., $\Omega>a_1\omega$) and the second term in the parenthesis of Eq.~\eqref{eq:mu_sst} becomes dominant over the first term,  which is the conventional eddy-viscosity formulation, $\mu_t = \rho k/\omega$, in the $k-\omega$ model. 

The set of nine parameters given in Table \ref{tab:SST_coef} comprise the unknown coefficients $\bm{c}$ that must be specified in the Menter SST model. Based on the sensitivity analysis by \citet{schaefer_uncertainty_2017} and the calibration step described in Section \ref{subsec:calibrate}, we will reduce the number of unknown parameters to only four: $\beta^*$, $\beta_1/\beta^*$, $\beta_2/\beta^*$, and $a_1$. These are the parameters that will be estimated in the present demonstration of ABC-IMCMC. 

\subsection{Axisymmetric transonic bump\label{sec:overflow_solver}}
\begin{figure}[!t]
	\centering
	\subfigure[Axisymmetric transonic bump experiment]{\includegraphics[width=0.5\linewidth]{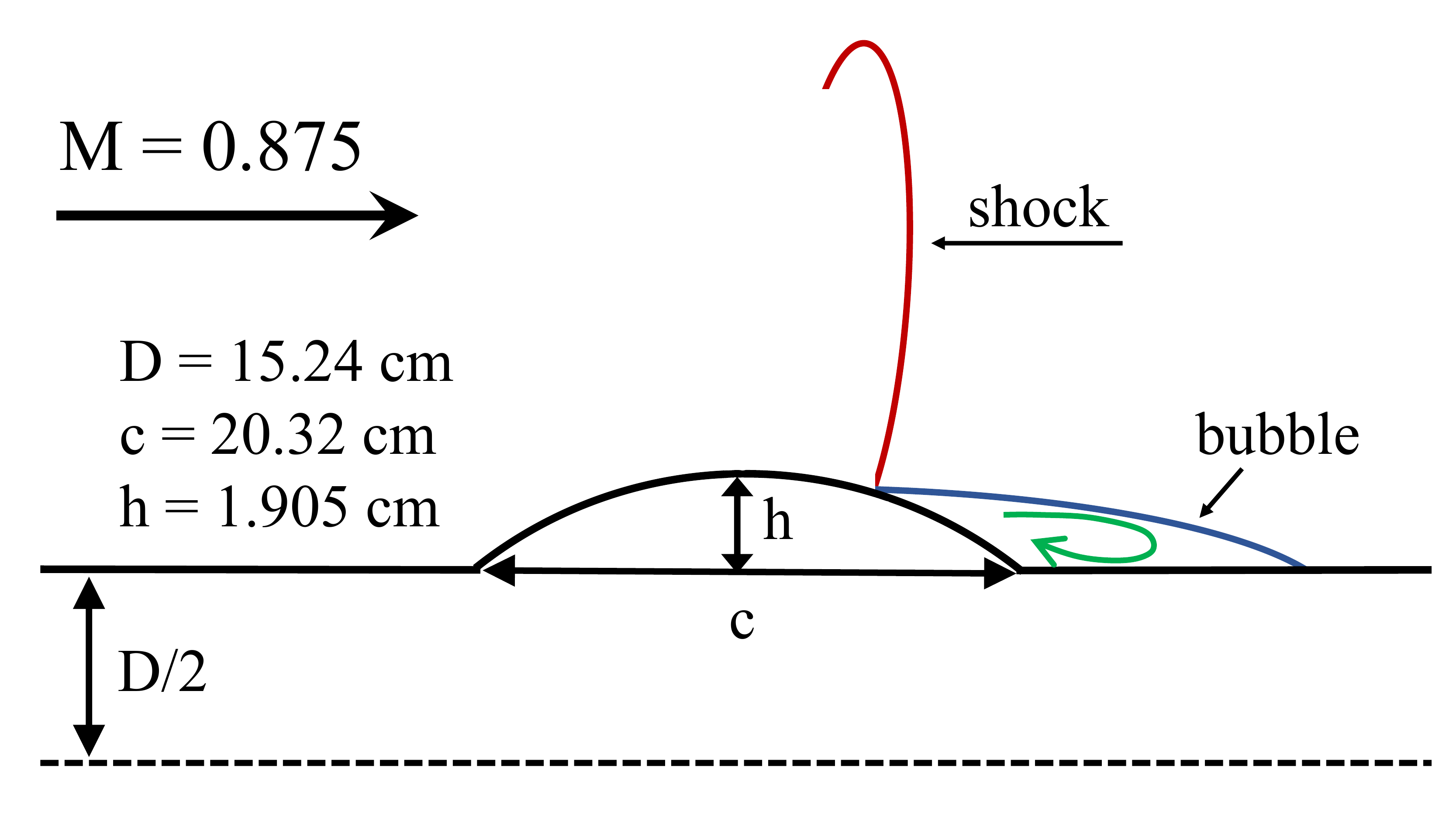}\label{fig:overflow_test_parameters}}
	\hfill
	\subfigure[Axisymmetric transonic bump simulation]{\includegraphics[width=0.45\linewidth]{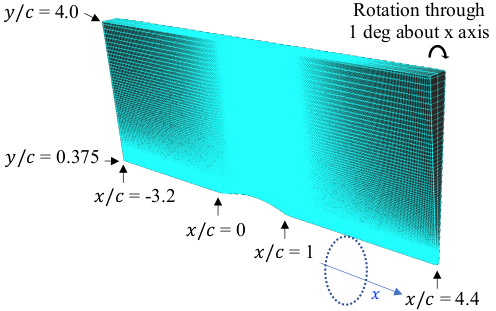}\label{fig:overflow_test_geometry}}
	\caption{Schematics showing (a) the experimental setup of the axisymmetric transonic bump geometry, including dimensions and operating conditions \cite{bachalo1986transonic}, and (b) the computational mesh used in the axisymmetric RANS simulations with the Menter SST model in \texttt{OVERFLOW}.}
\end{figure}

To estimate parameters in the Menter SST model, we use experimental reference data for an axisymmetric transonic bump from \citet{bachalo1986transonic}. This is a widely used test case for shock-induced separated flow, followed by reattachment. The axisymmetric bump in this experiment is a circular-arc bump with a height of $h=1.905$~cm and a length of $c=20.32$~cm attached to a cylinder that is $D=15.24$~cm in diameter. The flow has a freestream Mach number $M=U_{\inf}/a_{\inf} = 0.875$, where $U_{\inf}$ is the freestream velocity and $a_{\inf}$ is the speed of sound corresponding to a temperature of $T_{\inf} = 540^{\circ}$R. The combination of the shock and trailing-edge adverse-pressure-gradient causes flow separation with reattachment downstream of the bump (thereby creating a separation bubble).  The Reynolds number $\mathrm{Re} = 2.763\times 10^6$ is calculated based on $U_{\inf}$ and the bump length $c$.  All of the computational case parameters are provided on the TMR website\footnote{\url{https://turbmodels.larc.nasa.gov/axibump_val.html}\label{fnote:bump}} and Fig.~\ref{fig:overflow_test_parameters} shows a schematic of the experimental setup that is modeled in the simulations.  

Solutions of the Menter SST model for different parameters $\bm{c}$ were obtained here using the NASA \verb OVERFLOW ~code version 2.2n\footnote{\url{https://overflow.larc.nasa.gov/}}. \verb OVERFLOW ~is a 3D compressible flow code that solves the time-dependent RANS equations using multiple overset structured grids. The code can also  operate in two-dimensional or axisymmetric mode. It has been broadly verified and validated~\cite{childs2014overflow,jespersen2016overflow} and is widely used across the aerospace industry. 

The axisymmetric transonic bump case has been used previously for Menter SST model validation in \verb OVERFLOW ~\cite{jespersen2016overflow} and is thus ideally suited for the calibration of SST model parameters in the present work. Uncertainty quantification for the Menter SST model coefficients~\citep{schaefer_uncertainty_2017} has also been performed for this case using the same computational grid as that used here. Minor modifications were made to \verb OVERFLOW ~in order to allow the code to read turbulence model coefficients from the parameters file in order to perform the ABC-IMCMC algorithm. 

For axisymmetric problems, \verb OVERFLOW  can be configured in an axisymmetric mode with a three-plane grid. The center plane lies on the $x,z$ plane and the other two planes lie at $\pm1^\circ$ rotations from this plane. All computational solutions were obtained using a computational grid with  $721\times321$ cells on the center plane provided by the TMR website. The geometry and grid configuration are shown in Fig.~\ref{fig:overflow_test_geometry}. Each simulation ran for 5000 time steps, requiring $\sim 7$ minutes on a single node with two eight-core Intel Xeon E5-2670 2.6 GHz processors.

\subsection{Calibration and setup of ABC-IMCMC\label{subsec:calibrate}}
Prior to performing ABC-IMCMC, we complete a series of calibration steps that are intended to reduce the computational cost of the parameter estimation procedure, as well as provide the threshold $\epsilon$ and the initial values of $\bm{c}$ and the covariance of the proposal $q$ for the Markov chains. During the calibration process, we also select appropriate summary statistics for use with ABC-IMCMC. It is anticipated that this initial calibration will be used with any implementation of ABC that requires computationally expensive CFD simulations. To a large extent, this process also removes the need for substantial user input prior to initiating the Markov chains, since the selection of initial parameter values, covariances, and thresholds can be automated during the calibration. 

To begin, we immediately reduce the number of targeted unknown parameters in the Menter SST model from nine (i.e., those listed in Table \ref{tab:SST_coef}) to five. In particular, in their sensitivity study of the parameters in the Menter SST model, \citet{schaefer_uncertainty_2017} showed that $\beta^*$, $\sigma_{\omega1}$, $\beta^*/\beta_1$, $\beta^*/\beta_2$, and $a_1$ are the most sensitive parameters in simulations of the same axisymmetric transonic bump case examined here. Based on this analysis, we thus restrict the parameter estimation procedure to, at most, these five parameters, and the nominal values from Table \ref{tab:SST_coef} are used for the other four parameters.

The first calibration step is then subsequently used to determine appropriate summary statistics for the ABC-IMCMC procedure. For this step, we sampled $N=6^5=7,776$ parameter values, $\bm{c}$, from a five-dimensional uniform grid with 6 samples per dimension. Given the relatively small number of samples, we use the less computationally efficient classic ABC rejection algorithm (i.e., Algorithm~\ref{alg:abc-rej} in Appendix \ref{appendix}) for this calibration. The advantage of this algorithm during the calibration process is that the samples chosen do not depend on the acceptance criteria (as opposed to Markov chains, where samples are chosen based on knowledge of the previously accepted sample), allowing us to store the model output and measure the distance between reference and modeled summary statistics after all simulations have been performed. This allows us to build posterior distributions for various summary statistics and threshold values, and to choose the best statistics and thresholds for the parameter estimation using Markov chains. 

With respect to the available reference summary statistics, we use experimental data for the axisymmetric transonic bump provided by \citet{bachalo1986transonic}. The experiment provides the pressure coefficient $C_p=(\overline{p}-p_{\inf})/\frac{1}{2}\rho U_{\inf}^2$ along the bump wall, as well as eight mean velocity and turbulent shear stress transverse (i.e., wall normal) profiles at $x/c=-0.25$, 0.688, 0.813, 0.938, 1.0, 1.125, 1.25, and 1.375. Thus, the available reference summary statistics, which we denote here as $\mathcal{S}_k(x_i,y_j)$ for different streamwise, $x_i$, and wall normal, $y_j$, experimental measurement locations, are the pressure coefficient $\mathcal{S}_1(x_i,y_j)=C_p(x_i,y_j)$, the mean velocity $\mathcal{S}_2(x_i,y_j)=\overline{u}(x_i,y_j)/U_{\inf}$, the turbulent shear stress $\mathcal{S}_3(x_i,y_j)=\overline{u'v'}(x_i,y_j)/U_{\inf}^2$, and the separation and reattachment points for the separation bubble created after the leading shock, denoted $x_\mathrm{sep}$ and $x_\mathrm{reattach}$, respectively. Corresponding modeled summary statistics, $\mathcal{S}'_k(x_i,y_j)$, are then obtained from simulations in \texttt{OVERFLOW}, where the mean pressure and velocity are directly output by the code and the turbulent shear stress was calculated during post-processing as
\begin{equation}
    \overline{u'v'} = -\frac{\mu_t}{{\rho}}\left(\frac{\partial \overline{u}}{\partial y}+\frac{\partial \overline{v}}{\partial x}\right),
\end{equation}
where the turbulent eddy viscosity, $\mu_t$, was taken from SST model calculation. 

Given a particular summary statistic $\mathcal{S}_k(x_i,y_j)$, we define the distance function as the 2-norm of the difference between the reference and modeled statistics, namely
\begin{equation}\label{eq:dk}
    d_k(\mathcal{S}_k, \mathcal{S}'_k) = \left\{\sum_j \sum_{i} \left[\mathcal{S}'_k(x_i,y_j) - \mathcal{S}_k(x_i,y_j)\right]^2\right\}^{1/2},
\end{equation}
where summation over repeated indices is not implied, and $i$ and $j$ span the data points where the reference summary statistics are provided. To combine multiple types of reference data into a single distance function, we normalize each individual summary statistic distance, $d_k$ from Eq.~\eqref{eq:dk}, by the maximum over all $x_i$ and $y_j$, denoted $\max \mathcal{S}_k$, and by number of points $N_k$ in each summary statistic, yielding
\begin{equation}
    d(\mathcal{S}, \mathcal{S}') = \left\{\sum_{k}\frac{1}{N_k} \left[\frac{d_k(\mathcal{S}_k,\mathcal{S}_k')}{\max \mathcal{S}_k}\right]^2\right\}^{1/2}\,.
\end{equation}
For distance functions including error in the separation and reattachment points, we add as a condition that the corresponding statistical distance, $d_\mathrm{sep}$, must be less than a given threshold, where the distance is defined as
\begin{equation}
    d_\mathrm{sep} = \left[(x'_\mathrm{sep} - x_\mathrm{sep})^2 + (x'_\mathrm{reattach} - x_\mathrm{reattach})^2\right]^{1/2}\,.
\end{equation}
Here, $x_\mathrm{sep}$ and $x'_\mathrm{sep}$ indicate the reference and modeled separation points, respectively, and $x_\mathrm{reattach}$ and $x'_\mathrm{reattach}$ are the corresponding reattachment points.

\begin{figure}[t!]
	\centering
    \includegraphics{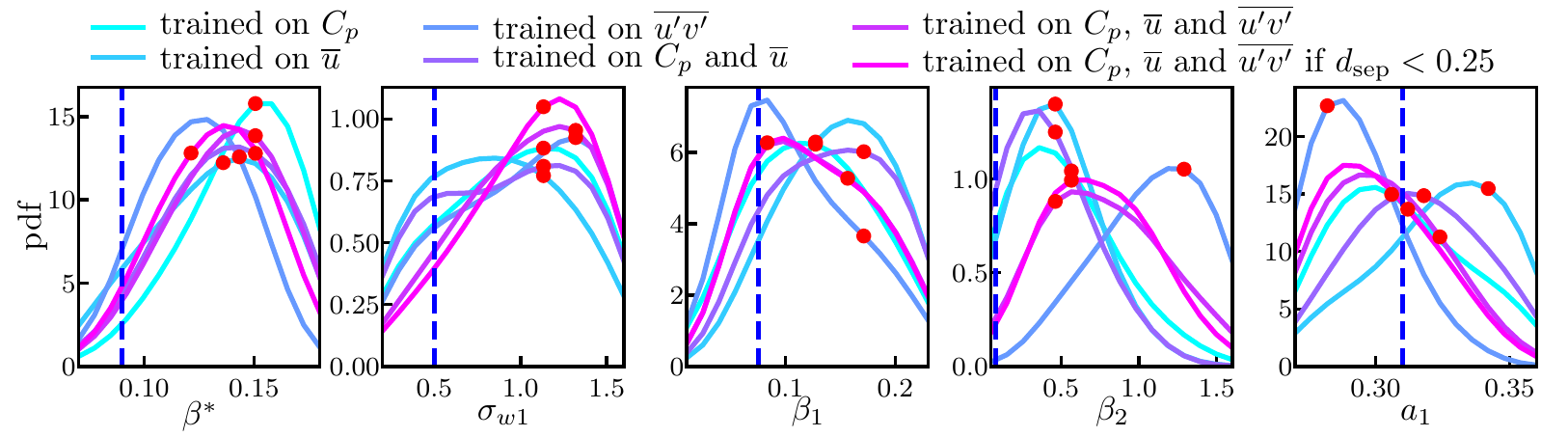}
	\caption{Marginal posteriors from the calibration of ABC-IMCMC for the Menter SST model and the axisymmetric transonic bump, showing results for $\beta^*$, $\sigma_{\omega1}$, $\beta_1$, $\beta_2$, and $a_1$, using a threshold that gives 3\% accepted parameters for 7,776 samples. Line colors correspond to the different summary statistics. Red dots show the MAP parameter values from the full five-dimensional posterior, and the vertical blue dashed lines show the nominal parameter values listed in Table~\ref{tab:SST_coef}.}
	\label{fig:5_params}
\end{figure}

Figure~\ref{fig:5_params} shows marginal posteriors from the ABC rejection algorithm for the five parameters $\beta^*$, $\sigma_{\omega1}$, $\beta_1$, $\beta_2$, and $a_1$ using the three different summary statistics individually (i.e., pressure coefficient, mean velocity, and turbulent shear stress profiles). This figure shows that the marginal posteriors for $\beta_2$ and $a_1$ have completely different MAP values when the shear stresses are used for the summary statistic in the ABC approach. This indicates that the SST model may not be able to simultaneously predict $C_p$, $\overline{u}$, and $\overline{u'v'}$ with a high degree of precision using the same parameter values.   

Figure~\ref{fig:5_params} also shows marginal posterior distributions after combining the summary statistics in various ways. In particular, we consider the following combinations: (\textit{i}) pressure coefficient, $\mathcal{S}_1$, and mean velocity, $\mathcal{S}_2$, (\textit{ii}) pressure coefficient, mean velocity, and turbulent shear stress, $\mathcal{S}_1$--$\mathcal{S}_3$, and (\textit{iii}) the same as (\textit{ii}), but with the condition that the separation and reattachment distance, $d_\mathrm{sep}$, be less than 0.25. Figure~\ref{fig:5_params} shows that the resulting marginals have maxima that are mixtures of the maxima from the marginals of each of the summary statistics individually.

\begin{figure}[!t]
	\centering
    \includegraphics{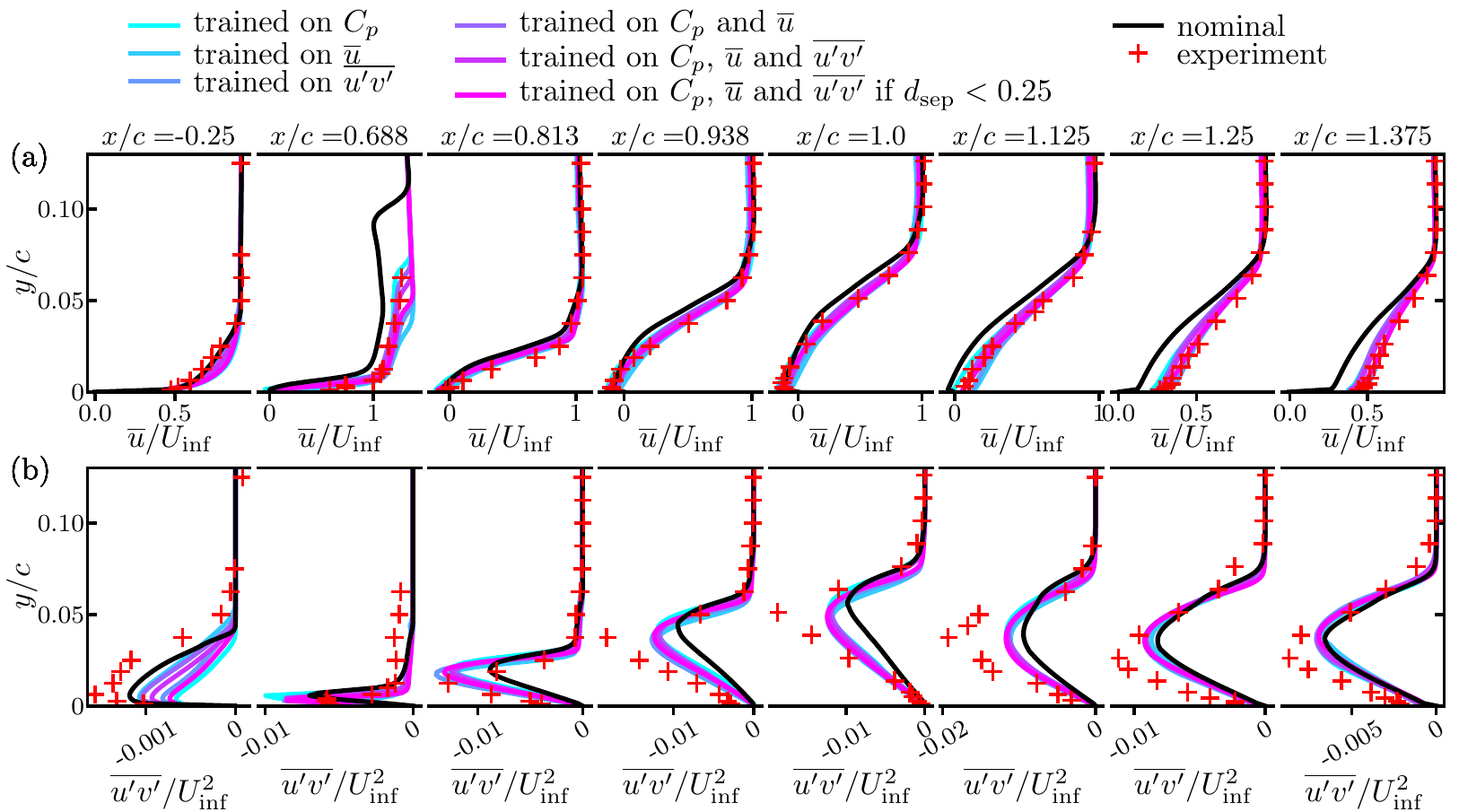}
	\caption{Experimental data \cite{bachalo1986transonic} and simulation results for (a) the mean velocity and (b) the turbulent shear stress produced using the Menter SST model with nominal coefficients from Table~\ref{tab:SST_coef} and the estimated MAP values from Fig.~\ref{fig:5_params}.}
	\label{fig:u_uv_map_5params}
\end{figure}

\begin{figure}[!t]
	\centering
    \includegraphics{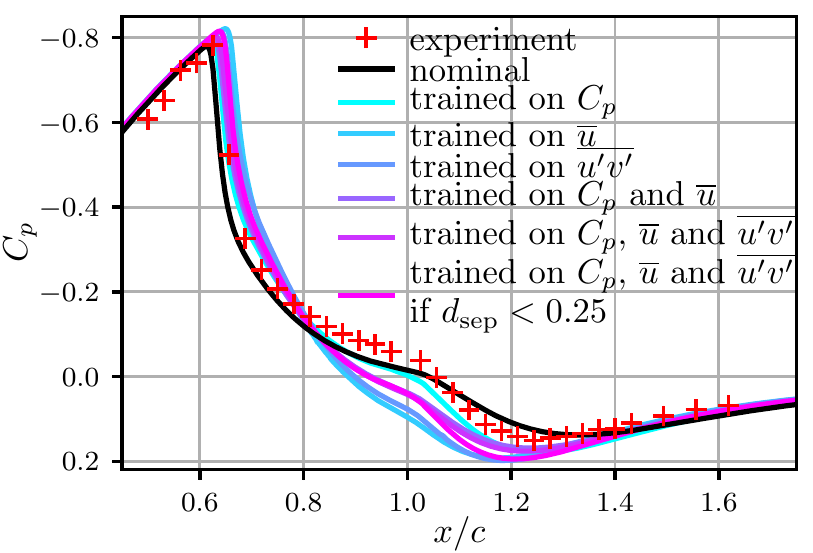}
	\caption{Experimental data \cite{bachalo1986transonic} and simulation results for the pressure coefficient $C_p$ produced using the Menter SST model with nominal coefficients from Table~\ref{tab:SST_coef} and the estimated MAP values from Fig.~\ref{fig:5_params}.}
	\label{fig:cp_map_5params}
\end{figure}

To compare results for different summary statistics, we estimated MAP values of the joint probability distributions with 3\% accepted parameters for all six summary statistics presented in Fig.~\ref{fig:5_params}, and we performed forward simulations with these parameters in \texttt{OVERFLOW}. Figure~\ref{fig:u_uv_map_5params} shows the resulting mean velocity and turbulent shear stress profiles, and Fig.~\ref{fig:cp_map_5params} shows the corresponding pressure coefficient profile. In general, as compared to model results using the nominal parameter values from Table~\ref{tab:SST_coef}, the model results with estimated parameters from ABC agree more closely with the experimental results for mean velocity and shear stress at nearly all locations. The nominal results are only better than the ABC results for these two summary statistics at the first measurement location, $x/c=-0.25$. Figure \ref{fig:cp_map_5params} shows that results for the pressure coefficient are generally similar for each of the models using the estimated parameter values from ABC. However, the nominal values from Table~\ref{tab:SST_coef} do give reasonable agreement with the experimental measurements for this statistic, despite the more noticeable disagreement for $\overline{u}$ and $\overline{u'v'}$ shown in Fig.~\ref{fig:u_uv_map_5params}.


Taken together, Figs.~\ref{fig:u_uv_map_5params} and \ref{fig:cp_map_5params} indicate that any choice of summary statistics in the ABC procedure will generally yield parameter estimates that provide better overall agreement with the experiments than the nominal values given in Table \ref{tab:SST_coef}. As such, we thus choose to use all available reference statistics in the remaining tests, resulting in a combined distance function based on the pressure coefficient, mean velocity, and turbulent shear stress, with the condition that the error in the separation and reattachment points must be less than 0.25.

\begin{figure}[!t]
	\centering
    \subfigure[Five parameters and 28,804 samples.]{\includegraphics{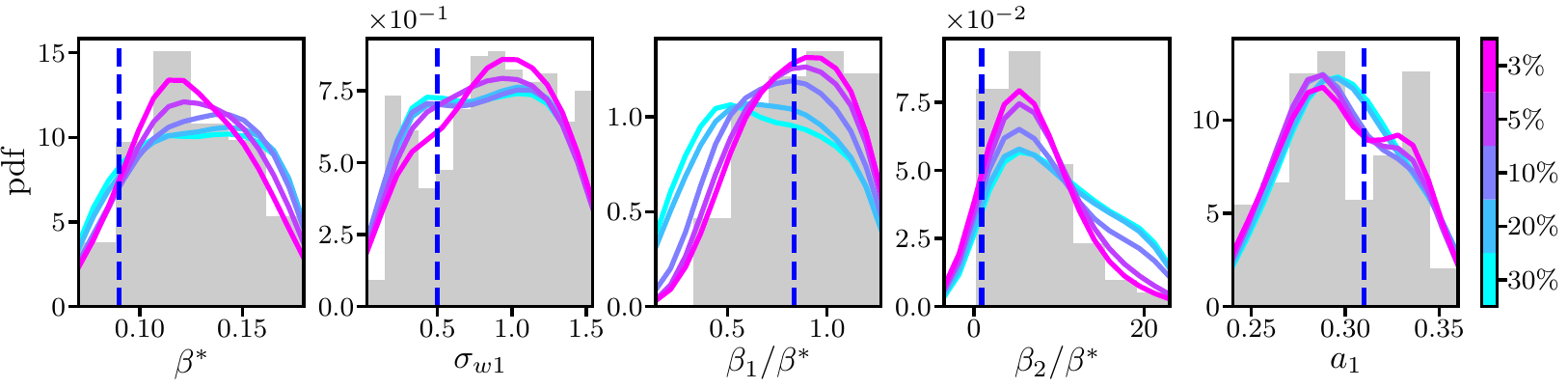}\label{fig:5_params_bb}}
    \subfigure[Four parameters and 20,736 samples.]{\includegraphics{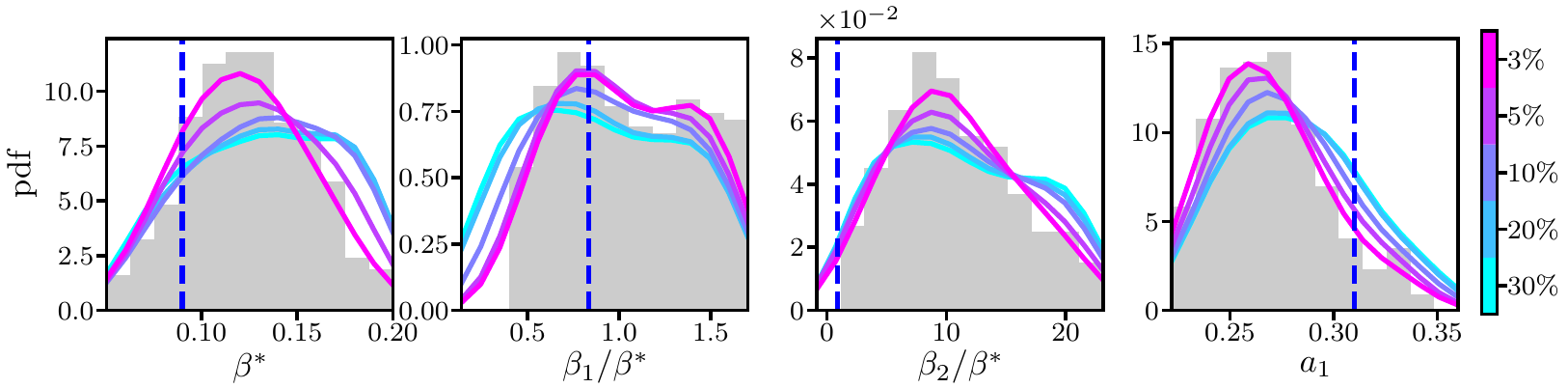}\label{fig:4_params_bb}}
	\caption{Marginal posteriors from the calibration of ABC-IMCMC for the Menter SST model and the axisymmetric transonic bump, showing results for (a) four and (b) five parameter calibrations. Line colors correspond to the percentage of accepted simulations, as determined by the rejection threshold $\epsilon$. Raw marginals for a 3\% acceptance rate (i.e., before Gaussian kernel density estimation) are shown by gray bars. Vertical blue dashed lines show the nominal parameter values listed in Table~\ref{tab:SST_coef}.}
\end{figure}

During the course of the first calibration step described above, we noticed a strong linear correlation between $\beta^*$ and $\beta_1$ in the posterior of accepted parameters. Correspondingly, we performed a second calibration step, again with the ABC rejection algorithm, with $N=28,804$ samples of the five parameters $\beta^*$, $\sigma_{\omega1}$, $\beta_1/\beta^*$, $\beta_2/ \beta^*$, and $a_1$ (i.e., a uniform grid with 6, 13, 6, 7, 8 samples per dimension, respectively). Figure~\ref{fig:5_params_bb} shows the resulting marginal distributions, where the larger number of samples for $\sigma_{\omega1}$ was chosen to demonstrate that the marginal distribution of this parameter is close to uniform. Thus, to reduce the amount of computation, we removed $\sigma_{\omega1}$ from the set of unknown parameters and set its value to the nominal value $\sigma_{\omega1} = 0.5$.

With these four parameters and the combined summary statistic, a final calibration step was performed with $N=12^4=20,736$ samples of the parameters $\beta^*$, $\beta_1/\beta^*$, $\beta_2/ \beta^*$, and $a_1$ (i.e., a uniform grid with 12 samples per dimension). Figure~\ref{fig:4_params_bb} shows the resulting marginal distributions for this step, revealing that the MAP values for each parameter are contained within the bounds of the parameter values, and that the posteriors are converging as $\epsilon$ decreases.

Using the four-dimensional posterior distribution from this final calibration, we defined necessary characteristics for the MCMC component of the ABC-IMCMC algorithm (see the schematic in Fig.~\ref{fig:algorithm}). In particular, we set the threshold $\epsilon$ such that $P[d(\mathcal{S}', \mathcal{S})\le\epsilon]=0.03$ (i.e., the maximum distance value of 3\% accepted parameters) and we defined the standard deviation of the initial Gaussian kernel for the Markov chains as 0.25 the standard deviation of the marginal posteriors formed by accepting 3\% of the sampled parameters. The starting parameter values for the chains were also randomly chosen from these accepted parameters.

\subsection{Results and discussion\label{sec:overflow_results}}
After completion of the calibration described in the previous section, we initiated 200 independent Markov chains for the estimation of $\beta^*$, $\beta_1/\beta^*$, $\beta_2/ \beta^*$, and $a_1$ in the Menter SST model, using all available experimental reference summary statistics for the axisymmetric transonic bump. Each of the chains were initialized using parameter values randomly chosen from the accepted parameters (with 3\% acceptance rate) in the final calibration step. For an initial period of $k = 100$ steps, the chains were progressed without kernel adaptation. 

\begin{figure}[!t]
	\centering
    \includegraphics{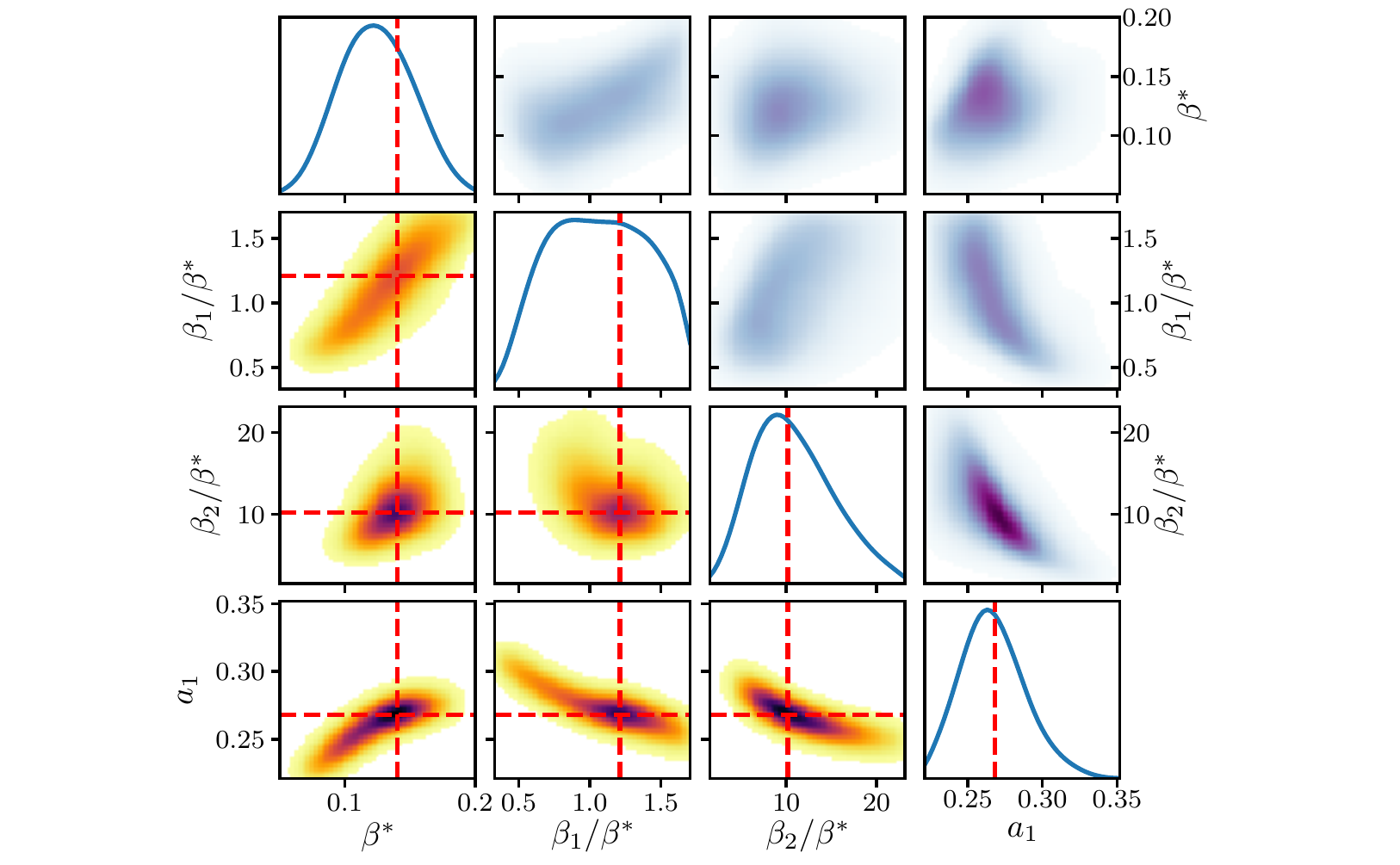}
	\caption{Posterior distributions of accepted parameters for the Menter SST model using experimental reference data for an axisymmetric transonic bump \cite{bachalo1986transonic}. Diagonal subplots show one-dimensional marginal pdfs, upper-diagonal subplots show 2D marginalized pdfs, and lower-diagonal subplots show conditional pdfs taken at the MAP values. The red dashed lines represent estimated parameters from the MAP values of the joint posterior.}
	\label{fig:chain_results}
\end{figure}

The Markov chains were then advanced, with kernel adaptation, according to Algorithm 2 in Appendix A (see also Fig.~\ref{fig:algorithm}). These chains yielded 118,052 accepted parameters that were then used to estimate the four-dimensional posterior distribution of unknown parameter values in the Menter SST model. The total number of sampled parameters (number of \verb OVERFLOW  runs) was 801,413 and required roughly 88,000 hours on a single node with two eight-core Intel Xeon E5-2670 2.6 GHz processors. The low acceptance rate was caused by the strict condition on the error in the separation and reattachment points, $d_{\mathrm{sep}} \le 0.25$.  This resulting posterior is the primary product of the ABC-IMCMC procedure, and Fig.~\ref{fig:chain_results} shows the resulting 1D and 2D marginalizations of the posterior, as well as the 2D conditional distributions at the MAP values of each pair of parameters. 

\begin{table}[t!]
	\centering
	\caption{Estimated parameters from ABC-IMCMC for the Menter SST model using experimental reference data for an axisymmetric transonic bump \cite{bachalo1986transonic}.}
	\label{tab:estimated_coef}
	\begin{tabular}{l|ccccc}
		\hline
		Coefficient   & $\beta^*$ & $\beta_1/\beta^*$ & $\beta_2/\beta^*$ & $a_1$ \\ \hline
		Nominal value & 0.09       & 0.833         & 0.92         & 0.31  \\
		Estimated values  & 0.14   & 1.212         & 10.17        & 0.268 \\
		\hline
	\end{tabular}
\end{table}

The MAP values of the posterior distribution shown in Fig.~\ref{fig:chain_results} provide estimates for the four unknown parameters; a summary of these values, along with the corresponding nominal values, is shown in Table~\ref{tab:estimated_coef}. Although the MAP estimates for $\beta^*$, $\beta_1/\beta^*$, and $a_1$ are different than the nominal values, they nevertheless remain the same order of magnitude. By contrast, the MAP estimate for $\beta_2/\beta^*$ is over ten times larger than the nominal value. This parameter is responsible for setting the balance of the production and dissipation of $\omega$, particularly further from the wall where the $k-\varepsilon$ model is active. This, in turn, corresponds to the location at which the shear stress in Fig.~\ref{fig:chain_results} is largest, and the ABC-IMCMC effectively increases $\beta_2/\beta^*$ to the greatest extent possible (i.e., without affecting the accuracy of other summary statistics) in order to drive the shear stress to larger magnitudes. This is also indicated by the calibration result shown in Fig.~\ref{fig:5_params}, where the posterior based on stress summary statistic corresponds to a substantially larger MAP value of $\beta_2$, as compared to the other cases.  

\begin{figure}[!t]
	\centering
    \includegraphics{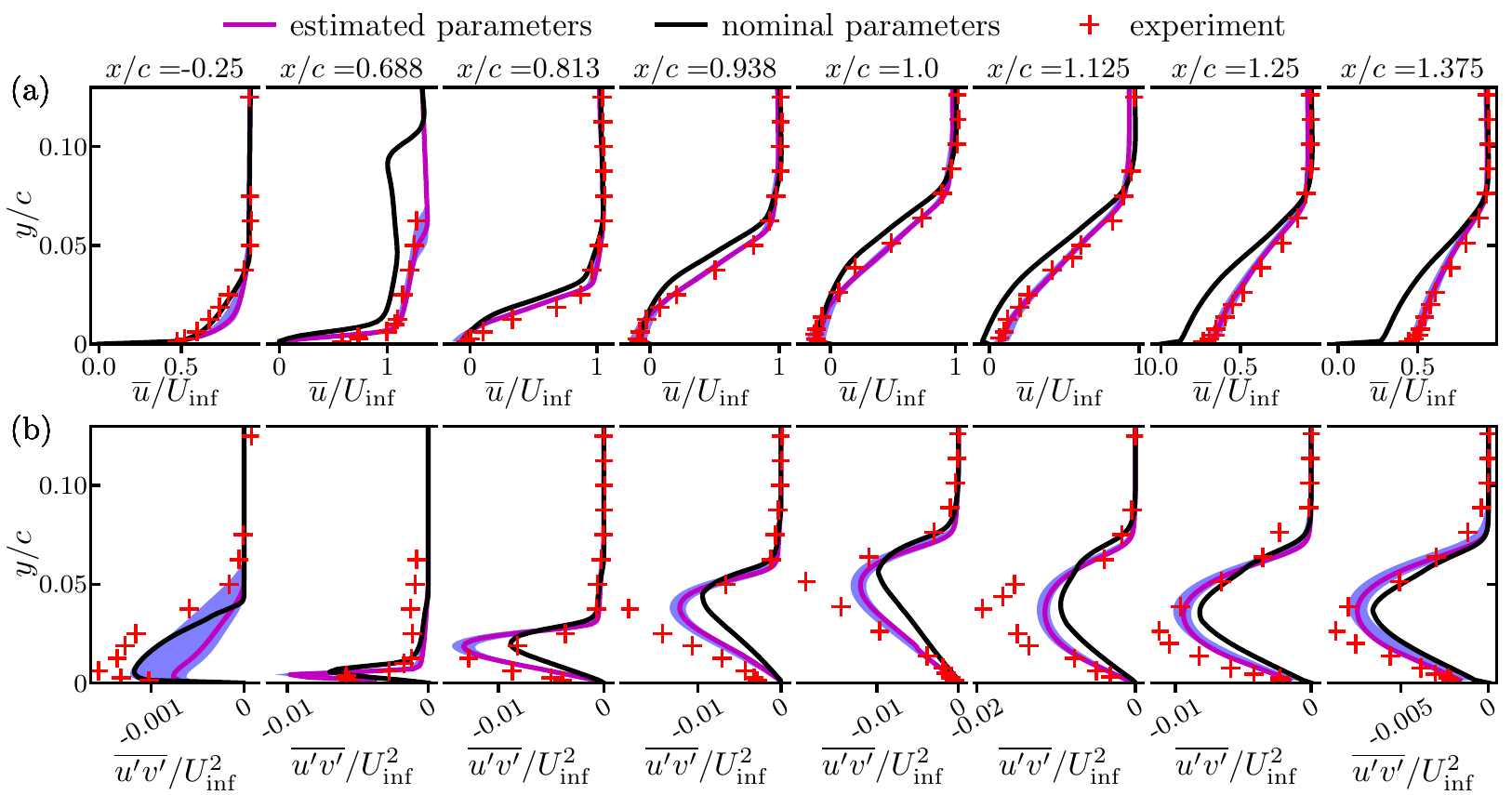}
	\caption{Experimental data \cite{bachalo1986transonic} and simulation results for (a) the mean velocity and (b) the turbulent shear stress produced using the Menter SST model with nominal coefficients from Table~\ref{tab:SST_coef} and the estimated MAP values from Table~\ref{tab:estimated_coef}. The shaded areas indicate the 95\% confidence interval.}
	\label{fig:u_uv_map_chains}
\end{figure}
\begin{figure}[!t]
	\centering
    \subfigure[Pressure coefficient $C_p$]{\includegraphics[width=0.49\linewidth]{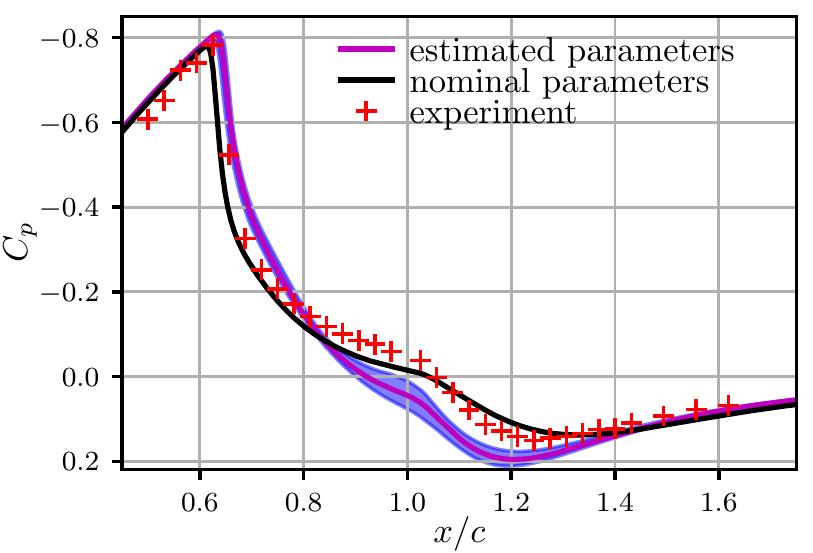}}
    \subfigure[Mean velocity along the surface]{\includegraphics[width=0.49\linewidth]{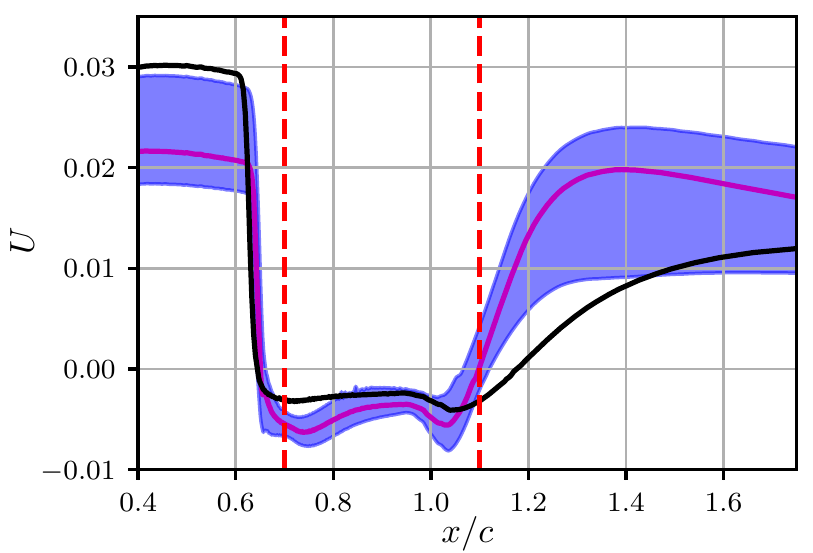}}
	\caption{Experimental data \cite{bachalo1986transonic} and simulation results for (a) the pressure coefficient $C_p$, and (b) the mean surface velocity produced using the Menter SST model with nominal coefficients from Table~\ref{tab:SST_coef} (black solid line) and the estimated MAP values from Table~\ref{tab:estimated_coef} (magenta solid line). The experimental data are shown as red crosses for the pressure coefficient and red dashed lines for the separation and reattachment points.  The shaded area indicates the 95\% confidence interval.}
	\label{fig:cp_map_chains}
\end{figure}

Mean velocity and turbulent shear stress profiles resulting from \texttt{OVERFLOW} simulations with the estimated and nominal parameter values are shown in Fig.~\ref{fig:u_uv_map_chains}, and Fig.~\ref{fig:cp_map_chains} shows the corresponding results for $C_p$ and the mean velocity along the wall (indicative of the separation and reattachment points). Figures~\ref{fig:u_uv_map_chains} and \ref{fig:cp_map_chains} show that there is a substantial overall improvement in the ability of the SST model to predict the experimental mean velocity and shear stress profiles, at the expense of slightly reduced agreement with the experimental measurements of $C_p$. However, the flexibility of the ABC-IMCMC method allows a user to easily adjust the weighting of different terms in the combined distance function such that improved agreement for $C_p$, or improvements in only a certain region of the flow (e.g., prior to flow separation), can be prioritized. The prediction of the separation point shown in Fig.~\ref{fig:cp_map_chains}(b) is also relatively similar for the estimated and nominal values, although the reattachment point is better predicted using the estimated parameters.

Beyond indicating the MAP values of the parameters, the posteriors in Fig.~\ref{fig:chain_results} also indicate that $a_1$ is positively correlated with $\beta^*$, and the widths of the posteriors indicate the degree of confidence in the MAP values as the ``best'' parameter estimates. Confidence in the MAP value of $\beta_1/\beta^*$ is relatively weak, for example, as indicated by the broader posteriors for this parameter. 

The effect of uncertainty in the MAP estimates can be propagated through the models to estimate uncertainty intervals on the output quantities of interest. To this end, we use all of the samples accepted in the ABC-IMCMC algorithm and estimate the distribution of the modeled profiles. The shaded areas in Figs.~\ref{fig:u_uv_map_chains}~and~\ref{fig:cp_map_chains} indicate the 95\% confidence interval for each distribution based on this sampling approach. In general, there is relatively little variability in the results for different choices of accepted parameters, and the greatest variability is observed in the mean velocity along the surface shown in Fig.~\ref{fig:cp_map_chains}(b). This variability along the surface is most likely connected to the broad posteriors for $\beta_1/\beta^*$ shown in Fig.~\ref{fig:chain_results}, since this parameter affects the balance of production and dissipation of $\omega$ in the near wall region, where the results in Fig.~\ref{fig:cp_map_chains}(b) are computed.

\section{Conclusions}
In this study, we have demonstrated the estimation of RANS turbulence model parameters using ABC. Many recent studies of model parameter estimation and uncertainty quantification have focused on statistical approaches. For example, Bayesian inference can be used to provide posterior distributions of unknown parameters, but knowledge of the likelihood function is required, which can be expensive to compute. By contrast, ABC uses a series of approximations to estimate the posterior without knowledge of the likelihood function.  Applying ABC to turbulence model parameter estimation reduces the computational burden of Bayesian inference and provides a flexible tool for model calibration.

Here we have provided a detailed description of the ABC methodology, including the baseline algorithm, acceleration using MCMC sampling, a calibration step, and an adaptive proposal. To demonstrate the use of ABC for turbulence model calibration, we estimated model parameters in a nonequilibrium RANS model applied to periodically sheared homogeneous turbulence, and in the Menter SST model applied to an axisymmetric transonic bump. Through these demonstrations, we showed the ABC-IMCMC approach to be an effective and efficient method for estimating unknown model parameters, as well as their uncertainties.

Although the MCMC procedure and adaptive proposal accelerate the ABC process and reduce the requirement for computational resources during the parameter estimation, the choice of summary statistics is a crucial component of ABC. These statistics must be sufficiently sensitive to changes in model parameters and must represent the dependence of the underlying reference data on these parameters. From this perspective, the classical ABC rejection algorithm, despite its computational expensiveness, has a noticeable advantage. Since parameter samples in this algorithm are independent of each other, we can store the output from the simulations with each sampled parameter and experiment with different choices of the summary statistic, distance function metric, and tolerance, to see how this affects our results. Thus, as shown here, the calibration step in the ABC-IMCMC algorithm can provide not only hyper-parameters for initiating the Markov chains, but can also be used to design the best summary statistics and distance function.

The estimation of turbulence model parameters in this study shows that models are often unable to simultaneously match all reference data fields, and ABC provides different estimated parameters for different summary statistics constructed from available reference data. As such, ABC provides an additional understanding of the model behavior and its ability to reconstruct real turbulent flows. Once the posterior distribution of model parameters has been estimated, we further showed that we can propagate parameter uncertainties through the model to estimate uncertainty intervals on the output quantity of interest.

Finally, it should be noted that the present study provides a demonstration of ABC for two types of RANS models and two different sets of reference data. However, the approach is completely general and can also easily be applied to other models and other flows. As such, the present demonstration should be taken as illustrative of the power of the ABC-IMCMC approach, and the framework has been uploaded to GitHub\footnote{\url{https://github.com/tesla-cu/turbABC}} for use by other researchers on other model development tasks in the future. The model parameters estimated here are also not intended to be taken as universally accurate; instead, the parameters are specific to the choice of model, choice of reference data, and particular setup of each ABC test. However, this does not preclude the development of more widely accurate models in the future; the flexibility of ABC easily allows a multitude of reference data sources and types to be used in the parameter estimation process, thereby increasing the generality of the models developed. 

\section*{Funding Sources}
OAD and PEH acknowledge financial support from NASA award NNX15AU24A-03. PEH was also supported, in part, by AFOSR award FA9550-17-1-0144. 

\section*{Acknowledgements}
Helpful discussions with Profs.\ Werner J.A.~Dahm, Ian Grooms, Will Kleiber, and Greg Rieker, as well as with Dr.~Jason Christopher, are gratefully acknowledged.

\newpage

\appendix

\section{Algorithms\label{appendix}}
\begin{algorithm}[H]
	\caption{ABC rejection sampling algorithm}\label{alg:abc-rej}
	\begin{algorithmic}[1]
		\State Calculate reference summary statistic $\mathcal{S}$ from $\mathcal{D}$
		\State Sample $N$ parameters $\bm{c}_i$ from prior distribution $\pi(\bm{c})$
		\For{each $\bm{c}_i$ }
		\State Calculate $\mathcal{D'} = \mathcal{F}(\bm{c}_i)$ from model
		\State Calculate model summary statistic $\mathcal{S}'$ from $\mathcal{D}'$
		\State Calculate statistical distance $d(\mathcal{S}', \mathcal{S})$
		\If{$d(\mathcal{S}', \mathcal{S})\le\epsilon$} 
		\State Accept $\bm{c}_i$
		\EndIf
		\EndFor
		\State Using all accepted $\bm{c}_i$ calculate posterior joint pdf
	\end{algorithmic}
\end{algorithm}

\begin{algorithm}[H]
	\caption{ABC-IMCMC algorithm with an initial calibration step and an adaptive proposal}\label{alg:IMCMC}
	\begin{algorithmic}[1]
	    \State Calculate reference summary statistic $\mathcal{S}$ from $\mathcal{D}$
		\Procedure{Calibration step}{$N_\mathrm{c}$, $r$}
		\State Sample $N_\mathrm{c}$ parameters $\bm{c}_i$ from prior distribution $\pi(\bm{c})$
		\For{each $\bm{c}_i$ }
		\State Calculate $\mathcal{D}^{\prime} = \mathcal{F}(\bm{c}_i)$ from model
		\State Calculate model summary statistic $\mathcal{S}'$ from $\mathcal{D}'$
		\State Calculate statistical distance $d_i(\mathcal{S}', \mathcal{S})$
		\EndFor
		\State Using all $d_i(\mathcal{S}', \mathcal{S})$ calculate distribution $P(d)$
		\State Define tolerance $\epsilon$ such that $P(d\le\epsilon)=r$
		\State Randomly choose $\bm{c}_0$ from $\bm{c}_i$ parameters with $d\le\epsilon$
		\State Adjust prior based on variance of parameters with $d\le \epsilon$
		\State Calculate covariance $\mathcal{C}_0$ from parameters with $d\le\epsilon$
		\EndProcedure
		\Procedure{MCMC without likelihood}{$\bm{c}_0$, $\epsilon$, $\mathcal{C}_0$, $k$, $N$}
		\State Start with accepted parameters $\bm{c}_0$ and covariance $\mathcal{C}_0$ 
		\State $i:=0$
		\While{$i < N$}
		\State Sample $\bm{c}'$ from proposal $q(\bm{c}_i\rightarrow \bm{c}')=q(\bm{c'}\,|\,\bm{c}_i, \mathcal{C}_i)$
		\State Calculate $\mathcal{D}^{\prime} = \mathcal{F}(\bm{c}')$ from model
		\State Calculate model summary statistic $\mathcal{S}'$ from $\mathcal{D}'$
		\State Calculate statistical distance $d(\mathcal{S}', \mathcal{S})$
		\If{$d(\mathcal{S}', \mathcal{S})\le \epsilon$}
		\State Accept $\bm{c}'$ with probability
		$$h = \min\left[1, \frac{\pi(\bm{c}')q(\bm{c}_i\rightarrow \bm{c}')}{\pi(\bm{c}_i)q(\bm{c}'\rightarrow \bm{c}_i)}\right]$$
		\If{Accepted}
		\State Increment $i$ 
		\State Set $\bm{c}_i = \bm{c}'$
		\State Update proposal covariance $\mathcal{C}_i$ as
		\[\mathcal{C}_i = 
        \begin{cases} 
        s_n \mathcal{C}_0, & \mbox{if } i < k\\ 
        s_n\mathrm{cov}(\bm{c}_0, \dots, \bm{c}_i), & \mbox{if } i\ge k 
        \end{cases}\]
        \EndIf
		\EndIf
		\EndWhile
		\State Using all accepted $\bm{c}_i$ calculate posterior joint pdf
		\EndProcedure
	\end{algorithmic}
\end{algorithm}

\end{document}